\documentclass[pdflatex,sn-mathphys-num]{sn-jnl}

\usepackage{graphicx}%
\usepackage{multirow}%
\usepackage{amsmath,amssymb,amsfonts}%
\usepackage{amsthm}%
\usepackage{mathrsfs}%
\usepackage[title]{appendix}%
\usepackage{xcolor}%
\usepackage{textcomp}%
\usepackage{manyfoot}%
\usepackage{booktabs}%
\usepackage{algorithm}%
\usepackage{algorithmicx}%
\usepackage{algpseudocode}%
\usepackage{listings}%

\usepackage{comment}
\usepackage{lipsum}

\begin{document}

\title[Article Title]{Real-Time Brain-Computer Interface Control of Walking Exoskeleton with Bilateral Sensory Feedback}


\author*[1]{\fnm{Jeffrey} \sur{Lim}}
\email{limj4@uci.edu}
\equalcont{These authors contributed equally to this work.}

\author*[1]{\fnm{Po T.} \sur{Wang}}
\email{ptwang@uci.edu}
\equalcont{These authors contributed equally to this work.}

\author[1,2]{\fnm{Won Joon} \sur{Sohn}}

\author[3]{\fnm{Derrick} \sur{Lin}}

\author[1]{\fnm{Shravan} \sur{Thaploo}}

\author[6]{\fnm{Luke} \sur{Bashford}}

\author[6]{\fnm{David A.} \sur{Bjanes}}

\author[4]{\fnm{Angelica} \sur{Nguyen}}

\author[4]{\fnm{Hui} \sur{Gong}}

\author[4]{\fnm{Michelle} \sur{Armacost}}

\author[4]{\fnm{Susan J.} \sur{Shaw}}

\author[6]{\fnm{Spencer} \sur{Kellis}}

\author[4,5]{\fnm{Brian} \sur{Lee}}

\author[4,5]{\fnm{Darrin} \sur{Lee}}

\author[1,7]{\fnm{Payam} \sur{Heydari}}

\author[6]{\fnm{Richard A.} \sur{Andersen}}

\author*[1,7]{\fnm{Zoran} \sur{Nenadic}}
\email{znenadic@uci.edu}

\author*[4,5]{\fnm{Charles Y.} \sur{Liu}}
\email{cliu@usc.edu}

\author*[3]{\fnm{An H.} \sur{Do}}
\email{and@uci.edu}

\affil[1]{\orgdiv{Department of Biomedical Engineering}, \orgname{University of California, Irvine (UCI)}, \orgaddress{\city{Irvine}, \state{CA}, \country{USA}}}

\affil[2]{Work done at UCI, currently at \orgname{Abbot Laboratories}, \orgaddress{\city{Plano}, \state{TX}, \country{USA}}}

\affil[3]{\orgdiv{Department of Neurology}, \orgname{UCI School of Medicine}, \orgaddress{\city{Orange}, \state{CA}, \country{USA}}}

\affil[4]{\orgdiv{Department of Neurology}, \orgname{Rancho Los Amigos National Rehabilitation Center}, \orgaddress{\city{Downey}, \state{CA}, \country{USA}}}

\affil[5]{\orgdiv{University of Southern California (USC) Neurorestoration Center}, \orgname{Keck School of Medicine of USC}, \orgaddress{\city{Los Angeles}, \state{CA}, \country{USA}}}

\affil[6]{\orgdiv{Division of Biology and Biological Engineering}, \orgname{Caltech}, \orgaddress{\city{Pasadena}, \state{CA}, \country{USA}}}

\affil[7]{\orgdiv{Department of Electrical Engineering and Computer Science}, \orgname{UCI}, \orgaddress{\city{Irvine}, \state{CA}, \country{USA}}}


\abstract{Invasive brain-computer interface (BCI) technology has demonstrated the possibility of restoring brain-controlled walking in paraplegic spinal cord injury patients. 
However, current implementations of BCI-controlled walking still have significant drawbacks. In particular, prior systems are unidirectional and lack sensory feedback for insensate patients, have suboptimal reliance on brain signals from the bilateral arm areas of the motor cortex, and depend on external systems for signal processing. 
Motivated by these shortcomings, this study is the first time a bidirectional brain-computer interface (BDBCI) has demonstrated the restoration of both brain-controlled walking and leg sensory feedback while utilizing the bilateral leg motor and sensory cortices. 
Here, a subject undergoing subdural electrocorticogram electrode implantation for epilepsy surgery evaluation leveraged the leg representation areas of the bilateral interhemispheric primary motor and sensory cortices to operate a BDBCI with high performance. 
Although electrode implantation in the interhemispheric region is uncommon, electrodes can be safely implanted in this region to access rich leg motor information and deliver bilateral leg sensory feedback. Finally, we demonstrated that all BDBCI operations can be executed on a dedicated, portable embedded system. These results indicate that BDBCIs can potentially provide brain-controlled ambulation and artificial leg sensation to people with paraplegia after spinal cord injury in a manner that emulates full-implantability and is untethered from any external systems.}


\maketitle

\section{Introduction}\label{sec:intro}

Individuals with paraplegia due to spinal cord injury (SCI) typically experience both a loss of lower extremity motor and sensory function. This often results in severe gait impairment and a dependence on wheelchairs, which increases the risk of other comorbidities such as heart disease, osteoporosis, and/or pressure ulcers~\cite{johnson_cost_1996}. These individuals typically rate the recovery of ambulatory function highly among their rehabilitation priorities~\cite{anderson_targeting_2004,collinger_functional_2013}. Physical therapy is the current clinical standard of care and provides limited recovery~\cite{johnson_cost_1996}. 
Recently, spinal cord stimulation (SCS)~\cite{edgerton2011epidural} 
has demonstrated a dramatic ability to restore motor, sensory, and autonomic function in people with SCI.
However, this is typically only achieved in SCI recipients with more preserved function (i.e., ASIA C/D). More severely affected SCI recipients (i.e., ASIA A/B) have a $<$5\% chance of regaining independent walking~\cite{wankr:24,kandharis:22,rejce:17,gillm:18,wagnerf:18,rowalda:22}. The combination of an electrocorticography(ECoG)-based brain-computer interface (BCI) and SCS~\cite{lorach_walking_2023} provided some additional motor improvement over SCS alone, but was only tested in a single subject with significant retained motor function (ASIA C). As such, improved methods to restore gait and sensory function after SCI are still needed, especially for those with the most severe injuries.

Brain-computer interface (BCI) technology is one such method to restoring gait after SCI. BCIs can bypass the site of SCI by utilizing signals from the primary motor cortex (M1) to control devices for ambulation. This has been achieved by both non-invasive \cite{king_brain-computer_2014,do_brain-computer_2013} and invasive BCIs \cite{benabid_exoskeleton_2019}. 
Notably, the motor aspect of walking after SCI has received significant attention in BCI research and has overshadowed the restoration of sensation. However, such motor restoration without sensory function is highly suboptimal given the critical role of sensation in normal gait.
To address this, a bidirectional BCIs (BDBCI) for ambulation could potentially elicit artificial leg sensation during BCI-controlled walking via direct cortical electrostimulation (DCES) of the primary somatosensory cortex (S1)~\cite{hiremath_human_2017,flesher_intracortical_2017,lee_engineering_2018}.
However, such BDBCIs have yet to be demonstrated in SCI subjects; only unidirectional, invasive, ECoG-based BCIs controlling a robotic gait exoskeleton (RGE) exist~\cite{benabid_exoskeleton_2019}.
Specifically, BDBCIs for walking with leg sensory feedback have not advanced beyond a highly limited feasibility demonstration in able-bodied individuals \cite{lim2024early}.

While the above examples indicate that the restoration of brain-controlled ambulation with leg sensory feedback in paraplegic SCI subjects is possible, the existing BCI and BDBCI technologies for walking still face several shortcomings.
First, in both of the ECoG-based BCI examples above \cite{benabid_exoskeleton_2019,lim2024early}, the decoding performances were inferior to that achieved by an EEG-based BCI for walking \cite{king_brain-computer_2014}.
This is likely attributed to the fact that both studies utilized signals from the lateral M1. This area predominantly represents the upper extremities and only has limited lower-extremity motor representation compared to the robust leg representation in the interhemispheric leg M1 area~\cite{mccrimmon_electrocorticographic_2018}. Although the interhemispheric leg area is difficult to access surgically, it is likely optimal for invasive BCIs for walking and allows the delivery of artificial leg sensation. 
Finally, existing BCI systems typically rely on external computers to perform computationally intensive tasks, e.g., motor decoding.
Such constant reliance wirelessly tethers the BCI to external systems and limits their practicality and energy efficiency in mobile applications. Instead, an embedded systems design approach where all functions are executed on board, would allow future BDBCIs to be highly compact, fully independent, and fully implantable systems. 
However, even though the BDBCI system in \cite{lim2024early} leveraged an embedded system approach to achieve leg percepts during BCI operation, it only involved unilateral sensory feedback.

Motivated by these outstanding problems, the current work demonstrates a BDBCI-RGE system for bilateral motor and sensory restoration of ambulation by utilizing bilateral interhemispheric ECoG implants.
Real-time, bilateral sensorimotor functions were achieved using an embedded systems approach that served as a benchtop analogue to a fully-implantable device. This functionality was safely demonstrated 3 weeks post-implantation in an able-bodied epilepsy subject with bilateral interhemispheric ECoG grids, and high performance was rapidly established over the course of 9 days.

\section{BDBCI-RGE System Overview}
\begin{figure}
    \centering
    \includegraphics[width = \linewidth]{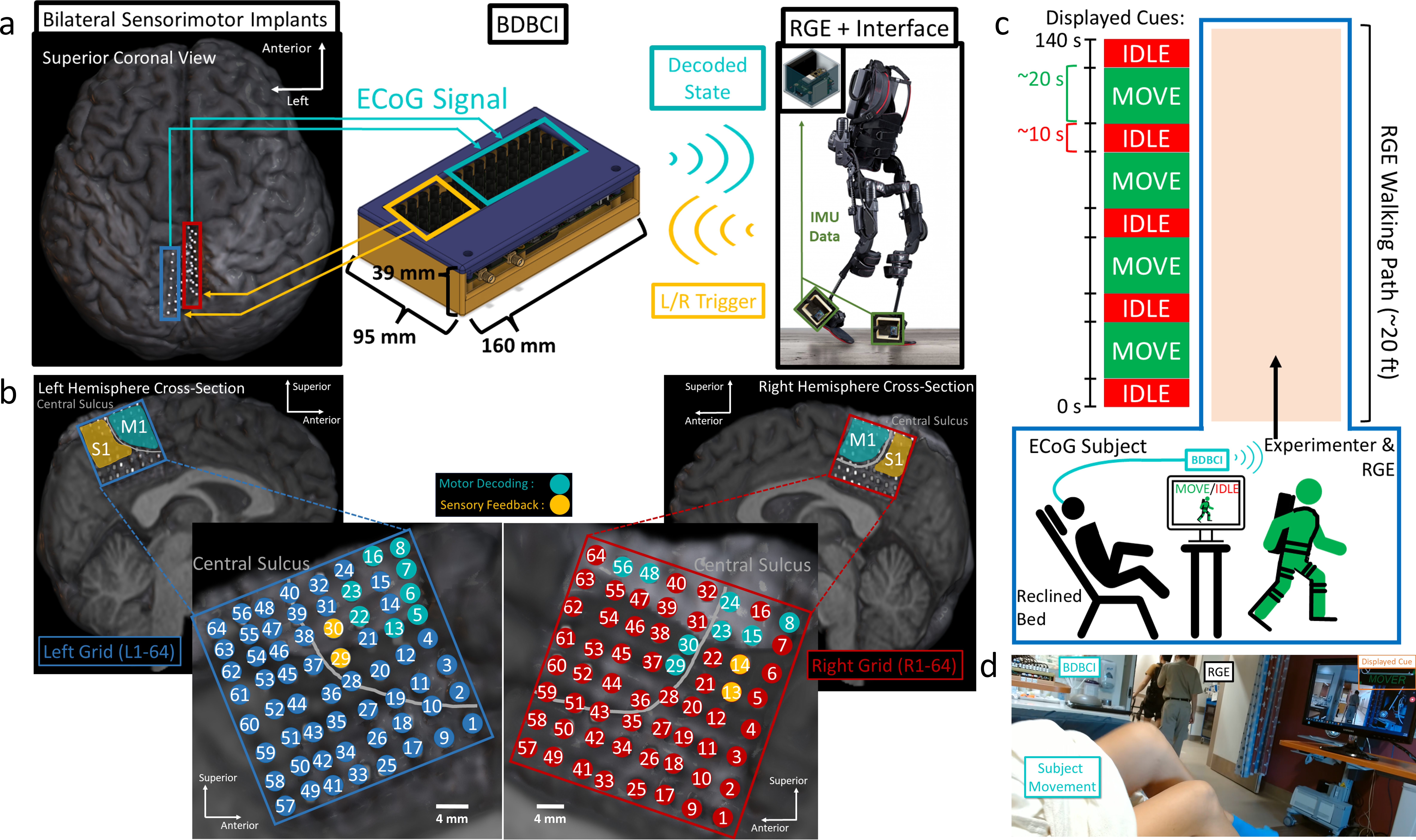}
    \caption{\textbf{a}: System diagram for BDBCI-RGE. ECoG signals are decoded to determine motor intent, which is then used to wirelessly control the RGE. RGE Interface detects right/left leg swing and commands the the BDBCI stimulator module to elicit the corresponding sensation using DCES.
    \textbf{b}: 
    Co-registration of ECoG electrode locations (post-implantation CT image) onto a 3D render of subject's brain (post-implantation MRI). Electrodes used for motor decoding on the left/right interhemispheric grids are colored teal. Electrodes used for sensory stimulation are colored yellow.
    \textbf{c}: Diagram summarizing BDBCI-RGE task. Subject seated in ICU bed (chair configuration) wirelessly controls the RGE by following cues to idle (relax body) or move (seated stepping motion with both legs). An experimenter wearing the RGE being walked down a pathway in response to BDBCI control, triggering sensory feedback (DCES) to the subject on each step. \textbf{d}: Photo of BDBCI-RGE task.}
    \label{fig:overview}
\end{figure}

Our system comprises the BDBCI device, RGE, and the RGE Interface (Fig.\ref{fig:overview} a). The BDBCI device, implemented on a custom printed circuit board (PCB), was designed to interface with common, commercially-available, ECoG electrode arrays, and contains both motor decoding and sensory DCES functions. 
It uses FCC Medical Device Radiocommunications Service (MedRadio) band wireless transceivers to communicate with the RGE Interface and a computer base station. Note that the computer base station is not necessary for online function and is only used to set online decoding parameters, control stimulation settings, and log experimental data. 
The RGE (Ekso GT, Ekso Bionics, Richmond, CA, USA) is an FDA-cleared powered exoskeleton for human ambulation after SCI.
The RGE Interface is mounted on the RGE and acts as a wireless interface between the BDBCI and the RGE, converting the BDBCI control signals to RGE step actuation and relaying the leg kinematic information to the BDBCI to trigger artificial sensory feedback.

\section{Subject Information and Implantation Procedure} \label{sec:subject_info}
 The study was approved by the Institutional Review Boards (IRB) at the University of California, Irvine and Rancho Los Amigos National Rehabilitation Center. An epilepsy patient (F, Age 50) undergoing Phase II epilepsy surgical evaluation at Rancho Los Amigos National Rehabilitation Center provided informed and written consent to participate in this study. The subject was implanted with two high-density 8$\times$8 grids (PMT Corporation, Chanhassen, MN, USA, 2 mm electrode diameter, 4 mm pitch) subdurally in the left and right interhemispheric fissures over the leg M1/S1 (electrodes L1--64 (R1--64) in the left (right) hemisphere, respectively). The location of these ECoG grids is shown in Fig.~\ref{fig:overview} b. Note that there were also additional ECoG electrodes not depicted in Fig. \ref{fig:overview} b, and all electrodes were placed for clinical epilepsy surgery evaluation. The ECoG electrodes were implanted for a total of 19 days, during which the subject stayed in the Intensive Care Unit (ICU). Study procedures were performed between Day 11 and 19. Additional details regarding the subject and the implantation procedure can be found in the supplement (Supplement, Section 1).

\section{Motor Mapping Procedure} \label{sec:motormapping}
\begin{figure}
    \centering
    \includegraphics[width = \linewidth]{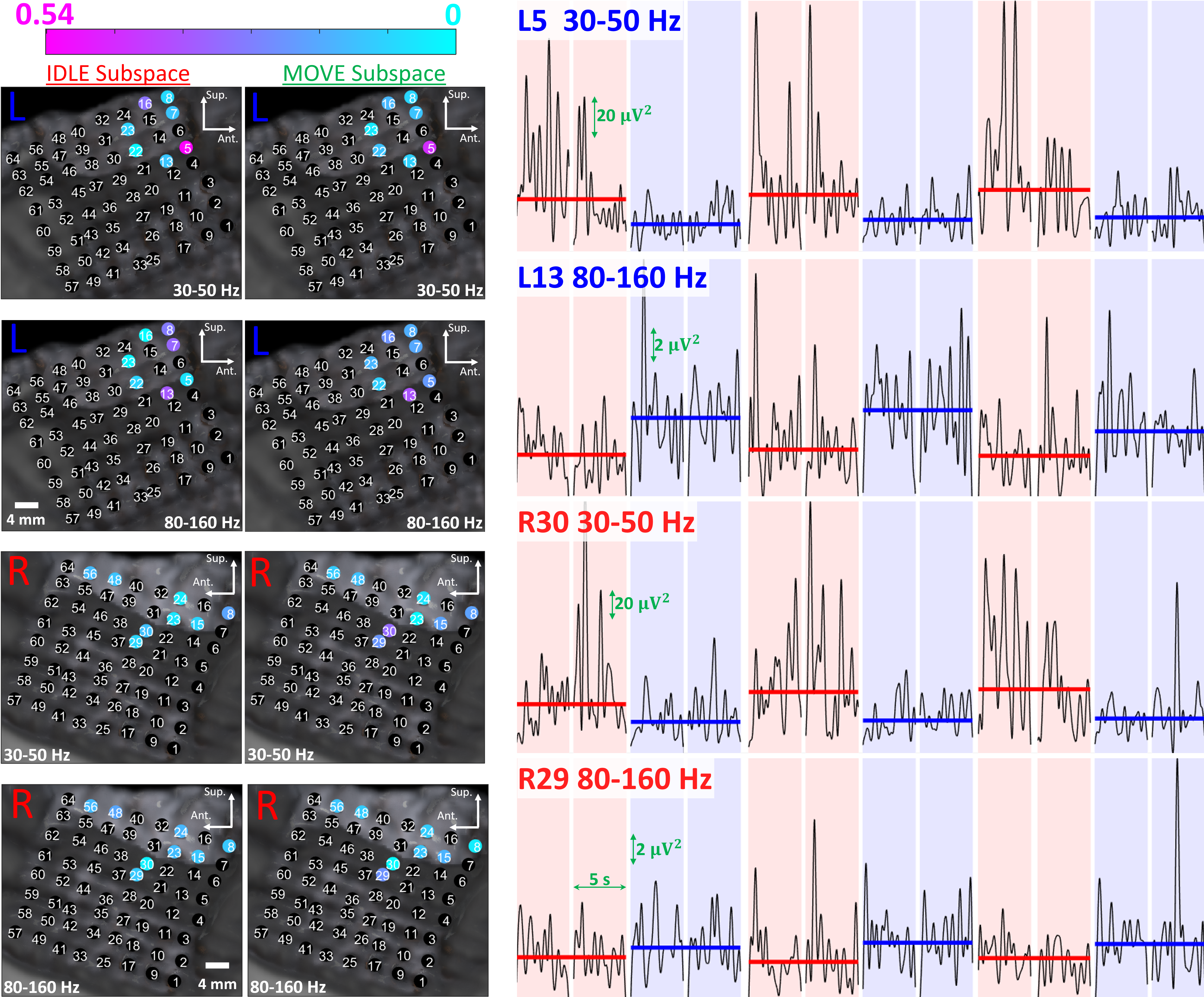}
    \caption{\textbf{Left:} Spatial weights of motor decoding model mapped to locations of corresponding coregistered ECoG electrodes in the left and right interhemispheric grids. Warmer colors indicate electrodes containing relevant decoding features for the given frequency band (high-$\beta$ or $\gamma$), whereas cooler colors indicate electrodes that were less relevant for decoding. \textbf{Right:} Time-domain power envelopes from a subset of training data in the electrodes with the strongest decoder weights (two electrodes are shown from each side). The $\gamma$ power increased during the ``MOVE'' state (i.e., transitioning from red shade to blue shade). Conversely, high-$\beta$ power decreased during the ``MOVE'' state (desynchronizations). \textit{Red and blue shades:} ``IDLE" and ``MOVE" states, respectively. \textit{Red and blue horizontal bars:} median band powers in each state. White gaps represent brief discontinuities between recording epochs, where no data were acquired.}
    \label{fig:training_summary}
\end{figure}

To facilitate brain-controlled walking, we first performed a motor mapping procedure to identify electrodes with ECoG signal modulation in response to lower extremity movement. The subject performed stepping leg movements in the ICU bed (configured to an upright seated position) while ECoG data were acquired by the clinical amplifier array (Natus\textsuperscript{\textregistered} Quantum$^\textnormal{TM}$, Natus Medical Incorporated, Pleasanton, CA, USA, 512 Hz sampling rate). 
These data were analyzed for modulation in common physiological EEG bands (see Section~\ref{sec:methods_motormap} for details). 
To optimize the BDBCI decoding performance, only a subset of 15 electrodes (Fig. \ref{fig:overview} c) exhibiting robust motor-related modulation in the high-$\beta$ (30--50 Hz) and $\gamma$ (80--160 Hz) bands (Fig. \ref{fig:training_summary}) were ultimately chosen for the BDBCI-RGE task. 

\section{Sensory Mapping Procedure} \label{sec:sensorymapping}
To enable sensory feedback during BDBCI operation, we first identified stimulation parameters (electrode pair, current amplitude, pulse frequency) that could elicit artificial sensation for the left and right legs. Specifically, we performed a cortical mapping procedure by systematically varying the stimulation parameters  
(100--300 Hz for pulse frequency, 1.61--10.86 mA for current amplitude) were delivered using the BDBCI's stimulator module. To reduce the mapping parameter space, the anodic/cathodic phase width was fixed at 250 $\mu$s and the total stimulation duration was maintained at 2 s, the approximate duration of an RGE-actuated step. The mapping space was also restricted to channels exhibiting sensory responses when stimulated during the functional cortical mapping performed by clinicians. These were primarily electrodes lying in the superior half on each grid (see Supplement Section 2, for more details). 
After each stimulation, the subject was asked to report the quality and location of the sensation elicited, if any.
The full set of responses can be found in supplement (Supplement Table 1 and 2). Ultimately, we chose the stimulation channel L29--30 (Fig. \ref{fig:overview} b), with current amplitude of 8.22 mA and pulse frequency of 100 Hz, eliciting a tingling sensation spanning the posterior aspect of the right lower leg to the heel.
For the left leg, we chose channel R13--14, with current amplitude of 3.80 mA and pulse frequency of 300 Hz, eliciting a sensation of tingling in the posterior aspect of the left lower leg. 

After the left/right leg stimulation parameters were chosen, the reliability of each percept was evaluated with two sensory tasks. First, a step counting task validated the subject's ability to identify discrete sensations and provided evidence that the responses from the sensory mapping were repeatable.
An experimenter (outside of the subject's vision and hearing) wearing an inertial measurement unit (IMU) walked between 2--8 steps for 14 trials. The experimenter's leg swing on these steps wirelessly triggered stimulation on a leg percept, and the subject counted the number of steps in each trial. This was repeated for both left and right leg percepts (total 28 trials). The subject correctly counted 14 of 14 trials for the right leg percept (L29--30) and 12 of 14 trials for the left leg percept (R13--14). These results were tested against Monte Carlo simulation and were found to be significant (empirical p-value $<$10$^{-6}$).

Second, the blind sensory identification task was designed to assess the subject's ability to discriminate between left and right sensations, as well as to ensure each sensation was distinguishable from a ``null" sensation (i.e., stimulation parameters that do not elicit percepts). A randomized sequence of stimulations (25 left/right/null each) was delivered. The subject was able to correctly identify the left leg, right leg, and null sensations at a rate of 84\%, 96\%, and 100\%, respectively. The subject's reported sequence was compared to Monte Carlo simulation and was found to be significant (empirical p-value $<$ 10$^{-6}$). For further details regarding the sensory tasks and results from additional subjects see Supplement, Section 3.

\section{Motor Decoding Model Training}\label{sec:training}
Prior to BDBCI-RGE operation, a model for decoding stepping intent from ECoG signals must be built to enable motor control. To this end, the BDBCI recorded the subject's ECoG signals during alternating periods of leg movement and idling. 
Specifically, a training protocol implemented onboard the BDBCI triggered the base station computer to present ``IDLE" and ``MOVE" instructional cues to the subject in alternating 10-s epochs for a total of 300 s (15 ``IDLE" and 15 ``MOVE" epochs). While seated, the subject was instructed to continuously perform stepping movements alternating with both legs when a ``MOVE" cue was displayed, and to relax and remain still when an ``IDLE" cue was displayed. Data collected during this protocol from the previously selected 15 channels (See Section~\ref{sec:subject_info}) were stored in memory on the BDBCI to build the decoder model (Section~\ref{sec:methods:training}). 
Although the BDBCI was capable of building the decoding model onboard (as in \cite{wang_benchtop_2019,sohn_benchtop_2022}), these training ECoG data were instead downloaded to the base station computer and analyzed to generate the decoding model. The base station computer's faster processing expedited the decoding model training, as time with the subject was limited (1-2 mins on computer v.s. 20-30 mins onboard).
ECoG high-$\beta$ and $\gamma$ band signals were used to generate the decoding model as they contained salient and localized features associated with stepping movements. 
For example, a robust high-$\beta$ band desynchronization was observed on L5 (left grid anterior-superior interhemispheric leg M1) and a robust $\gamma$ band synchronization was observed on R30 (right grid posterior-superior interhemispheric M1). 
Fig.~\ref{fig:training_summary} shows color-coded maps of decoder model spatial weights across both left and right ECoG grids as well as examples of the behavior on L5, R30 as well as other electrodes with salient modulation.
Once the decoding model was built, it was uploaded to the onboard memory for subsequent BDBCI online decoding operations.

\section{Online Familiarization}\label{sec:rt_control}
Before BDBCI-RGE operation, the decoding model was briefly validated independent of the RGE and stimulation to ensure proper function and familiarize the subject with online BCI control. To this end, the subject was instructed to follow alternating ``IDLE" and ``MOVE" instructional cues displayed on a screen similar to the procedure during decoder model training. ECoG signals were acquired by the BDBCI and were decoded in real time into the ``MOVE" or ``IDLE" states. These decoded states and the corresponding instructional cues were wirelessly transmitted to the base station computer for performance analysis. Online performance was evaluated by calculating the lag-optimized Pearson product-moment correlation coefficient between the cues and the decoded states ($\rho$). Here, the subject achieved an average online performance of $\rho=0.89\pm0.06$ with lag $2.2\pm0.4$ s over 5 runs ($\sim$4--5 min per run for 20--25 min total)

\section{BDBCI-RGE Operation}\label{sec:bdbci}
With both motor control and artificial sensory feedback established, a subject moves onto the online BDBCI-RGE task. An overview of the BDBCI-RGE task is illustrated in Fig.~\ref{fig:overview}c. Similar to the prior BCI procedures, the subject performed this task in bed in the seated position. A computer screen displayed ``IDLE"/``MOVE" cues and provided visual feedback through a real-time camera view of the RGE (donned by an experimenter) as it walked in response to decoded ECoG signals. Contralateral S1 sensory stimulation was enabled so that each RGE step triggered a percept for the corresponding leg using parameters identified in Section~\ref{sec:sensorymapping}. As explained in Section~\ref{sec:rt_control}, real-time motor control was performed by asking the subject to follow ``IDLE"/``MOVE" cues displayed in alternating 5-s epochs. During ``IDLE" cues, the subject was instructed to relax to keep the RGE still. During ``MOVE" cues, the subject was instructed to perform stepping movements with both legs to trigger BDBCI-RGE mediated walking. During online operation, if the BDBCI decoded an ``IDLE" state, the RGE remains still. While the BDBCI decoded a ``MOVE" state, the RGE walks the experimenter forward (nominal stepping rate of 0.5 Hz) and triggers the BDBCI to elicit leg percepts. Decoding and sensory feedback stimulation were interleaved alternately (see Methods for further details) to avoid interference with ECoG signals by stimulation artifacts. This prevents disruption of motor decoding given sensory feedback stimulation typically only occurs during RGE leg swings. Note that decoding was not necessary during leg swing, as the RGE's step mechanisms were pre-determined and could not be modified mid-swing.

\begin{figure}[!htpb]
    \centering
    \includegraphics[width=\linewidth]{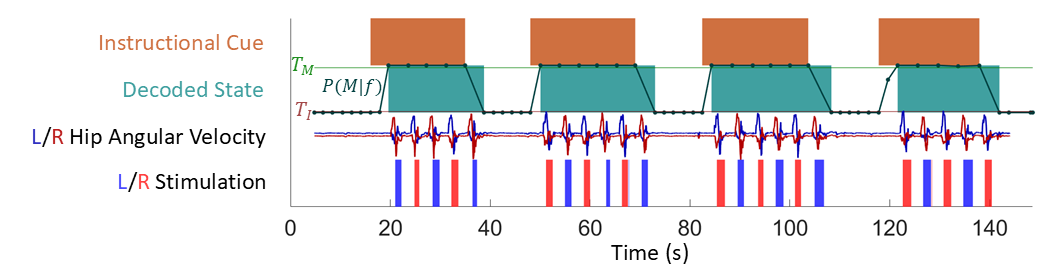}
    \caption{Best-case BDBCI-RGE online run. \textbf{Orange}: ``Instructional Cue" blocks indicate periods where the ``MOVE" cue was presented to the subject. \textbf{Teal}: ``Decoded State" blocks indicate when a ``MOVE" state was detected. The $P(M|f)$ line indicates the probability of ``MOVE", as determined by the decoder. Each dot on the line indicates the end of one online window. $T_M$ and $T_I$ indicate the state transition thresholds to which $P(M|f)$ was compared. 
    If the current decoded state is ``IDLE" and $P(M|f) > T_M$, the decoded state changes to ``MOVE". Conversely, if the current decoded state is ``MOVE" and $P(M|f) < T_I$, the decoded state changes to ``IDLE". 
    Once the decoded state is ``MOVE", the RGE begins to move, as indicated by the \textbf{L/R Hip Angular Velocity} from the motion tracking suit worn by the experimenter in the RGE. The \textbf{L/R stimulation} indicates the contralateral S1 stimulation corresponding to the movement of the matching RGE leg.}
    \label{fig:eksobdbci_online_run5}
\end{figure}

The subject performed 5 runs each day over two consecutive days  ($\sim$2.5 min per run, $\sim$25 min combined time). She achieved an average performance of $\rho=0.92\pm0.04$ with lag $3.5\pm0.5$ s across the 2 days ($\rho=0.89\pm0.02$ with lag $3.4\pm0.5$ s on Day 1 and $\rho=$0.94$\pm$0.03 with lag 3.6$\pm$0.5 s on Day 2).  Fig.~\ref{fig:eksobdbci_online_run5} illustrates the best result from one run of online BDBCI-RGE control. 
Note that for each cue change, the subject was able to transition the decoded state within 1 to 2 decoding windows (1.8 to 3.8 s) Even with these brief delays in reaction time, the decoding accuracy remained high.  
Anecdotally, the subject confirmed during all 10 runs that the RGE steps triggered the matching percept in the corresponding leg, and that this artificial sensation seemed to aid her in the task. See Supplement Section 4 for the results of individual runs.

To verify that BCI control was not driven by the effects of stimulation (i.e., decoder mistook electrical artifacts as motor cortex activation) or mirror motor neuron activity from observing the RGE, two control experiments were performed.
In the first control experiment (``BCI-RGE''), the RGE control was enabled with real-time camera feedback, but stimulation was disabled. 
In the second control experiment (``BCI-Stimulation''), the RGE control and the subject's camera feedback were turned off, and the BDBCI automatically stimulated left and right leg percepts (in 2-s alternation) when the ``MOVE'' state was decoded. 
Both conditions were performed for 5 runs, and the same cue scheme as the BDBCI-RGE task was followed (See Section~\ref{sec:rt_control}). The subject achieved an average decoding performance of $\rho = 0.93\pm0.01$ with lag $3.6\pm0.9$ s for the ``BCI-RGE'' control, and  $\rho = 0.94\pm0.02$ with lag $3.8\pm0.4$ s for the ``BCI-Stimulation'' control. These performances were comparable (two-sided Wilcoxon rank sum test, $p = 0.75$ and $p = 0.27$, respectively) to those achieved with BDBCI-RGE runs, indicating that neither mirror motor neuron activity nor the effects of stimulation primarily drove the decoding performance (c.f. Supplement Section 4 for results of individual control experiment runs).

No adverse events were reported during the course of this study. 
More specifically, no seizures were induced as a result of any of the cortical mapping, sensory tasks (Section~\ref{sec:sensorymapping}) or subsequent BDBCI procedures (Section~\ref{sec:bdbci}). The subject experienced pain percepts (felt like she stepped on something) for channels during clinical mapping (R52-53, R53-54, R55-56). These were deliberately avoided during the BDBCI sensory mapping. 

\section{Discussion}\label{sec:disc}

Herein we have described a BDBCI-RGE system that enabled real-time brain-controlled walking and provided bilateral leg sensory feedback. 
This was safely achieved in a subject with ECoG implantation in the interhemispheric leg M1 area, and with high decoding performance. 
Furthermore, sensory leg percepts were validated to be consistent and repeatable, and were confirmed to not interfere with the BDBCI performance. Finally, this was achieved with the entire BDBCI implemented on a portable embedded form factor without compromising overall function.
 
Whereas our prior work~\cite{lim2024early} demonstrated a BDBCI-RGE utilizing ECoG grids implanted on the lateral convexity M1, the current work represents the first implementation of a BDBCI for walking that incorporates bilateral inter-hemispheric leg sensorimotor areas. 
Our current work achieved a high performance of $\rho$ up to 0.97 online decoding accuracy (overall average $\rho$ of 0.92) over a short period of time ($\sim$9 days). This represents a marked decoding performance increase compared to \cite{lim2024early} ($\rho$ of 0.80). 
This difference in performance may be attributed to a more direct coverage of the leg M1 area in the interhemispheric space. The leg M1 area likely provides more robust and reliable localized ECoG modulation associated with leg movements. Furthermore, the BCI performance using signals from the lateral M1 convexity \cite{lim2024early,benabid_exoskeleton_2019} was inferior to that achieved by an EEG-based BCI \cite{king_brain-computer_2014}, which reinforces the notion that the interhemispheric region is likely the most optimal recording site for BCI-walking applications.
Finally, although interhemispheric ECoG implants are uncommon and complex to perform, this study demonstrates this location of implantation can be safely leveraged for invasive BDBCI systems.

Our BDBCI system is also the first to implement bilateral lower extremity artificial sensation alongside motor decoding in a single BDBCI-RGE system. By comparison, our work ~\cite{lim2024early} only involved unilateral sensory feedback. Beyond this prior report, there have been no other BDBCIs for gait and the lower extremities.   
The primary concern for employing DCES for sensory feedback is the risk of compromising the motor decoding, either through electrical artifacts~\cite{zhou_toward_2018,lim_artifact_2022} or evoked neural activity. We addressed these problems by interleaving motor decoding and stimulation processes. The results from our control experiments (Section~\ref{sec:bdbci}) suggest that the BDBCI operation was driven by M1 ECoG activity underlying leg movement intent, and not by stimulation artifacts, evoked motor neuron activity, or mirror neuron activation. 
This approach is similar to that used in~\cite{weissjm:18}, which employed amplifier blanking to accomplish interleaving of acquisition and stimulation processes. Such approaches are suboptimal as they introduce discontinuities in decoding and delay BDBCI responses. The normal human sensorimotor loop requires constant simultaneous sensory feedback, which is not permitted by interleaving or blanking. Hence, this makes it difficult for users to appreciate critical gait events such as transitions to double support, gait perturbations, etc. This continues to be a problem in BDBCI technologies, and will require novel methods of real-time stimulation artifact elimination to achieve ``full duplex'' capability.

Our BDBCI-RGE device achieved high performance with and without the artificial sensory feedback. Though BCI decoding performance showed no significant change with sensory feedback (as seen in the control experiments), it is possible that such feedback instead improves functional outcomes.
More specifically, artificial sensory feedback may facilitate improved functional and safety measures associated with BDBCI operation, such as faster gait velocity or reduced risk of falls by providing sensory confirmation of weight transfer and acceptance. 
Although not directly tested here, such a finding would be consistent with prior work \cite{flesher_intracortical_2017,flesher2021brain} which showed that sensory feedback can improve functional outcomes in upper extremities neuroprosthesis application.  

An important aspect of our BDBCI is that all functions are implemented as an embedded system within a compact footprint. While a base station computer was used, it was only required for setting parameters, initiating/aborting decoding functions, and logging experimental data. 
In particular, the BDBCI can operate entirely independently once configured and initialized via the base station computer. The portability of the entire system could be further improved by replacing the base station computer with a mobile device or tablet. 
Furthermore, the existing BDBCI form factor would allow it be readily mounted onto the RGE, and therefore enable applications where the subject (as opposed to an experimenter) is in the RGE.
By comparison, other existing BDBCI systems such as \cite{flesher_intracortical_2017,flesher2021brain}, rely on large, rack-mounted, off-the-shelf systems, or are constantly tethered to external computers for signal processing, which limit their applicability in mobile applications.
Although the current system is not yet implantable, the design lends itself to be translated into as a fully implantable device (discussed below). 
When compared to existing implantable BCI walking applications as \cite{benabid_exoskeleton_2019,lorach_walking_2023}, a fully implantable incarnation of our BDBCI is expected to operate fully independently on rechargeable battery power. As such, improved mobility, power efficiency, and aesthetics could be achieved by eliminating the need for constant wireless streaming to a permanently tethered computing system and constant magnetic power induction from an external power source.

A limitation of this study is that it was only performed in one able-bodied epilepsy subject. However, this subject provided a unique opportunity, as the bilateral interhemispheric ECoG implants used here for Phase 2 epilepsy surgery evaluations are extremely rare. An additional limitation is that the motor behavior used in this study was seated step-like movements instead of stepping while standing or actually walking. 
This leaves open the concern that the high BDBCI performance demonstrated here may not generalize to the paraplegic SCI population. For example, one concern in translating the BDBCI-RGE from able-bodied epilepsy subjects to SCI subjects is that ECoG modulation underlying walking intent may not be present or sufficient for BCI control. Though leg M1 behavior is generally assumed to be intact post-injury, it is possible that the neural circuitry has changed as a direct result of the injury or due to an extended period of disuse~\cite{cramer2005brain}. However, previous literature has indicated that this behavior can be retrained or reacquired by SCI subjects. For example, King \textit{et al.}~\cite{king_feasibility_2015} demonstrated the ability for a SCI subject to control an EEG-based BCI connected to lower-extremity FES. Although the subject initially had a low degree of control over the BCI due to inadequate EEG motor-related modulation, the subject was eventually able to achieve extremely high performance in BCI-controlled walking after 21 weeks.
Similarly, Benabid \textit{et al.}~\cite{benabid_exoskeleton_2019} 
and Lorach \textit{et al.}~\cite{lorach_walking_2023} demonstrate the ability for SCI subjects to operate BCI gait systems 1--2 years post-implantation using signals from the M1 lateral convexity  area. 
Future work will need to directly address these limitations by implanting the BDBCI device in the target SCI population. Anecdotally, our BDBCI-RGE system has received an FDA IDE to proceed with such an investigation.

Ultimately, we envision a fully implantable version of our device, which was previously described in~\cite{sohn_benchtop_2022}. 
Briefly, a skull-implanted unit interfaces with ECoG electrodes placed over M1/S1, and performs signal acquisition and stimulation channel selection. This skull unit is connected to a chest-wall implanted unit via a subcutaneous tunneling cable. The chest unit contains electronic components to perform all BDBCI functions, including signal processing, generation of electrical stimulation pulses, and wireless communication with base station and end effector units.
Such a fully-implantable BDBCI would facilitate chronic implantation in future paraplegic SCI recipients without the need for any transdermal components that could pose infection risks. 
The work presented here serves as an important foundation for the advancement of BDBCI technology towards such a fully implantable system, which may in turn present new avenues for treating sensorimotor deficits in affected patient populations.
However, miniaturization of the BDBCI system is necessary for fully-implantability.   
We expect the BDBCI to be implementable with a footprint comparable to that of existing implantable neural devices (e.g., NeuroPace RNS\textsuperscript{\textregistered} system 60 $\times$ 27.5 mm, Medtronic Percept\textsuperscript{\texttrademark} PC neurostimulator 68 $\times$ 51 mm). A unidirectional version of our system \cite{wang_benchtop_2019} already reached a similar size. Further applying advanced multilayer PCB design techniques, smaller surface-mounted components, and CMOS system-on-a-chip (SOC) integration of all electronic hardware components
can drastically reduce the device footprint. 
For example, our earlier work showed that significant footprint and power reduction was achieved via CMOS integrated circuit implementation of an amplifier array and stimulator module \cite{malekzadeh-arasteh_energy-efficient_2019,malekzadeh-arasteh_fully-integrated_2021,pu_40v_2021,pu_cmos_2021}.
However, more comprehensive SOC implementation and subsequent testing and validation are expensive and are therefore typically carried out only after the proof-of-concepts for the hardware are fully demonstrated. 

Additional future directions may include providing sensory training to enhance subject accuracy in sensory discrimination and detection in BDBCI operation. Other non-BDBCI sensory applications may also be implemented, such as skin ischemia-related warning percepts or bladder distension signals. Outside of neuroprosthesis applications, the BDBCI may also be able to utilize the neurorehabilitative effects of DCES. Inspired by prior work demonstrating that combining BCI and spinal cord stimulation provided rehabilitative benefits after SCI \cite{lorach_walking_2023}
the addition of sensory feedback to BCI-controlled spinal cord stimulation may activate novel Hebbian processes which could improve the recovery of function in damaged sensory pathways.

\section{Methods}\label{sec:methods}

\subsection{Study design and participants}
Adult (age$>$18) subjects undergoing Phase II epilepsy surgical evaluation with ECoG grids implanted over the M1 and S1 areas were recruited for this study. Exclusion criteria included the presence of any neurological impairment other than epilepsy, or medical condition that could prevent safe participation in the study. ECoG implantation and research procedures were carried out at Rancho Los Amigos National Rehabilitation Center. The sites of ECoG electrode implantation were based on clinical needs only. Study procedures were performed over the course of subjects' ICU stay. Any procedures requiring DCES were only performed while subjects were awaiting ECoG explantation surgery after clinical epilepsy monitoring was completed and anti-seizure medications resumed.

\subsection{CT-MRI Co-registration Procedure} \label{sec:methods_ctmri_coreg}
To identify the locations of the implanted electrodes relative to the brain anatomy, we co-registered the ECoG electrode locations from post-implantation head computed tomography (CT) with the brain tissue from magnetic resonance imaging (MRI) images. This would allow for an approximate location of those electrodes overlaying the M1/S1 leg area in the interhemispheric space. Co-registration was performed using a custom MATLAB (MathWorks Inc., Natick, MA, USA) script~\cite{wang_co-registration_2013,lim_artifact_2022} and the Elastix image processing toolbox~\cite{klein_elastix_2010,shamonin_fast_2014}. Post-operative CT and MRI images were used to maximize the accuracy of spatial localization.

\subsection{Motor Mapping} \label{sec:methods_motormap}
ECoG data recorded by the hospital's neural signal acquisition system (Natus\textsuperscript{\textregistered} Quantum$^\textnormal{TM}$, Natus Medical Incorporated) were analyzed to inform the selection of electrodes and frequency bands for decoding of seated stepping movements. Specifically, ECoG signals from all electrodes on both interhemispheric grids were recorded while a subject was verbally cued to alternate between the ``IDLE'' and ``MOVE'' tasks in $\sim$10-s intervals. During the ``IDLE'' cue, the subject was instructed to relax and avoid moving any parts of their body. During the ``MOVE'' cue, the subject was to perform alternating stepping motion with both legs while seated in bed. Both tasks were performed while the hospital bed was configured to an upright seated position. The ECoG data were exported to MATLAB. These data were segmented by the verbal cues, and the data from each channel were visually and quantitatively inspected to determine which channels exhibited physiological ECoG modulation in a preset list of  bands (i.e., $\mu$ (8--12 Hz), $\beta$ (12--25 Hz), high-$\beta$ (30--50 Hz) or $\gamma$ (80--160 Hz)) to inform channel and band selections for subsequent decoding model training (Section~\ref{sec:methods:training}).

\subsection{Sensory Mapping} \label{sec:methods_sensemap}
To facilitate selection of stimulation channel and parameters for leg sensory percepts, a mapping of the interhemispheric leg S1 was performed using the BDBCI's stimulator module. This process was informed by the results of the clinical cortical mapping procedures.
The clinical cortical mapping was performed by clinicians as part of surgical evaluation for epilepsy treatment to identify eloquent brain areas to spare from potential resection.
Briefly, the clinical cortical mapping procedures involved stimulating adjacent pairs of ECoG electrodes with a clinical stimulation system (Natus\textsuperscript{\tiny\textregistered} Nicolet\textsuperscript{\tiny\texttrademark} Cortical Stimulator, Natus Medical Incorporated) using a systematic escalation of stimulation current amplitudes (up to 20 mA). Other waveform parameters such as pulse frequency, train duration and pulse width were held constant, nominally at 50 Hz, 2 s and 250 $\mu$s, respectively. 
Channels and parameters that elicited a sensory response were noted, as they could potentially be used for sensory feedback.  
Cortical mapping using the BDBCI was performed similarly to the clinical mapping procedure, but the pulse frequency was also varied in the search space. The sensory percepts discovered in the clinical mapping procedure were re-verified using the BDBCI stimulator. The BDBCI was also used to search the expanded pulse frequency parameter space not reacheable by the clinical stimulation system.
Ultimately, one left and one right leg sensory percept elicited from the contralateral interhemispheric ECoG grid were chosen for use with the BDBCI. The reliability of these percepts were subsequently verified using the step counting (Section~\ref{sec:methods_stepcount}) and blind sensory (Section~\ref{sec:methods_blindsense}) experiments.

\subsection{Step Counting Experiment} \label{sec:methods_stepcount} 
The step counting experiment was performed to verify that the stimulation parameters identified in the previous mapping procedure could reliably elicit the corresponding percept, and that the percept could be used to discern individual steps. 
An experimenter wore an IMU device on the upper thigh. When they stepped with that leg, the IMU device wirelessly commanded the BDBCI to elicit a sensory percept on the chosen channel upon each step taken by the experimenter. The subject was tasked with identifying the number of steps using sensation alone. This was performed in 14 ten-second trials, wherein the experimenter would walk a predetermined number of steps (in another room, out of the subject's sight and hearing). The number of steps varied between 2 and 8 steps in each trial. For each trial, the subject would be prompted to count the number of discrete leg percepts they felt using a hand counter. The subject was not informed of their accuracy until after the entire test. 
This test was performed one leg percept at a time for both legs (28 trials total). The subject's performance for each leg percept was quantified by the number of trials correctly counted and was compared against 10$^6$ iterations of Monte Carlo (MC) simulation. 
Specifically, for each iteration, 14 ordered random integers were drawn from a uniform distribution $[2,8]$ and were compared to the number of steps taken by the experimenter, resulting in a correct number of matches $\in [0,14]$. The empirical p-value was the fraction of MC iterations that corresponded to the responses of the subjects.

\subsection{Blind Sensory Experiment} \label{sec:methods_blindsense}
The blind sensory experiment was performed to verify that subjects could discriminate between left leg, right leg and ``null'' percepts. A null percept was ``elicited'' by a set of stimulation parameters where no sensorimotor response was reported in the BDBCI sensory mapping. Using the same stimulation parameters identified above, a randomly permuted list of 25 of each percept was delivered to the subject, for a total of 75 stimulation events. 
For each stimulation, the subject was asked to identify whether a left leg, right leg, or null percept was delivered. The subject was informed when stimulation was delivered, but not which percept. The subject was also not informed of their accuracy until after the test. The subject's performance in this task was quantified by calculating the confusion matrix between the three percepts and deriving the proportion of correctly identified percepts. A MC simulation randomly permuted the list for 10$^6$ iterations and was compared to the subjects' responses. 
The empirical p-value was the fraction of MC iterations that matched the subjects' responses.

\subsection{BDBCI-RGE System Description}
\subsubsection{BDBCI Hardware} \label{sec:methods_hardware}
The BDBCI is implemented as a custom embedded system that interfaces with ECoG electrode grids via industry standard 1.5 mm touch-proof connectors. The board is housed inside a 3D printed case measuring 160$\times$95$\times$39 mm (Length, Width, Height). Three microcontroller cores (48 MHz, Microchip, Chandler, AZ, USA) executed all BDBCI functionalities, including signal acquisition, BCI decoding, stimulation control, setting stimulation parameters, and all wireless communications. A neural signal amplifier/digitizer chip (Intan Technology, Santa Monica, CA, USA) enabled acquisition for up to 32 channels (32 recording electrodes, 1 reference electrode, and 1 ground electrode) at 500 Hz and 16-bit resolution for training data recording or for real-time BCI operation. 
The programmable stimulator is implemented as a cascade of charge pumps with stimulation pulses generated by a programmable MOSFET-based H-bridge. It can generate biphasic square pulse train waveforms with current-controlled output up to 12 mA (controllable via a digital potentiometer), pulse frequency up to 300 Hz, 
and pulse widths up to 250 $\mu$s/phase \cite{sohn_benchtop_2022}. These limitations were chosen to keep the maximum delivery under the charge density safety limit (30 $\mu$C/cm$^{2}$/phase) \cite{Kuncel2004}. Furthermore, to prevent excessive charge build-up at the stimulation site, a charge-balancing protocol was incorporated as described in Sohn \textit{et al.}~\cite{sohn_benchtop_2022}. This protocol was facilitated by onboard current and voltage sensors.
Two transceivers operating at $\sim$406-MHz (FCC MedRadio band) enabled wireless communication with the base station computer and the RGE Interface (described below). System power was supplied by a 3.7 V, 2500 mAh lithium polymer battery. 

\subsubsection{Base Station Computer Software}
For a detailed explanation of the base station computer software for motor decoding and stimulation control, see Wang \textit{et al.}~\cite{wang_benchtop_2019} and Sohn \textit{et al.}~\cite{sohn_benchtop_2022}. Briefly, user interface (UI) software (implemented in C\# for the Microsoft Windows operating system) allowed experimenters to wirelessly configure and initialize the BDBCI. This includes configuring and starting the training data acquisition function (e.g., number of epochs, epoch duration) and online decoding function (e.g, state transition thresholds, online window duration), setting stimulation waveform parameters for left and right leg percepts (see Sec.~\ref{sec:methods_hardware}), as well as manual triggering of the stimulator. The UI program also receives and records real-time cues and decoder outputs from the BDBCI.

\subsubsection{RGE Interface}
The RGE Interface is a custom component that interfaced the BDBCI with the Ekso GT RGE. Its role is to trigger RGE stepping when commanded by the BDBCI and to command the BDBCI to elicit the corresponding left or right leg percept when RGE leg swing are detected. The RGE's hardware and software remained unmodified to maintain regulatory approval. 

The RGE Interface was implemented on a custom PCB consisting of a microcontroller connected to a wireless transceiver (MedRadio band), a microphone, a rotational servo motor attached to a cantilever, and two IMUs. 
The IMUs were mounted on the RGE ``ankles" to detect leg swings. A firmware-based binary state machine compared the IMU angular velocities to preset thresholds to determine the start and end of each RGE-mediated left or right leg swing. This information was wirelessly sent to the BDBCI to elicit the corresponding percept during leg swing. Additionally, each completed step would cause the RGE to emit an audio tone, which was captured by the microphone to inform the RGE that another step could be initiated safely.
The servo motor was attached to the RGE controller and depressed the step button with the cantilever upon command from the BDBCI. This setup avoids modifying the RGE, preserving all regulatory-approved safety features, and allows an experimenter to quickly disable the RGE interface if necessary.

\subsection{Decoder Training Procedure} \label{sec:methods:training}
Prior to real-time BDBCI operation, an online decoding model was generated using a supervised learning approach. To this end, ECoG data were acquired during alternating periods of ``MOVE"/``IDLE" behavior. Only channels containing salient leg movement related features during the motor mapping procedure were used (Section \ref{sec:methods_motormap}). A training data collection protocol on the BDBCI would then cue the subject to perform alternating epochs of move/idle behavior (as described in~\ref{sec:training}). Subjects were seated in the ICU bed (configured into an upright, seated position) and followed alternating 10-s ``IDLE'' or ``MOVE'' cues on the computer screen for a total of 300 s. The cues can be customized in each subject's preferred language. Subjects were instructed to relax and abstain from any movements when ``IDLE'' was displayed, and to perform a seated marching motion with both legs when ``MOVE'' was displayed. The BDBCI acquired and saved the ECoG signals in onboard memory. These stored signals were used to calculate the feature extraction matrices corresponding to the ``MOVE" and ``IDLE" states (classes). 

To calculate feature extraction matrices, the data were first common-average referenced. Each ``MOVE'' and ``IDLE'' trial was then subdivided into non-overlapping 750-ms windows. For each window and channel, the average powers in the selected bands were calculated. In most cases, features from two bands were sufficient to inform the difference between the two states, resulting in 
2 bands $\times$ 32 channels $\times\sim$48 samples per class. The 64-dimensional data underwent dimensionality reduction using class-wise principal component analysis (cPCA)~\cite{das_efficient_2009} for each class. Linear discriminant analysis (LDA) was then applied to enhance the separability between the two classes~\cite{fisher_use_1936,duda_hart_pattern_2000} and further reduced the data to a one-dimensional feature, $f$. The combined cPCA and LDA feature extraction matrices and other relevant parameters for each class were saved in the BDBCI onboard memory for use in real-time decoding.

\subsection{Decoder Architecture} \label{sec:methods_decoding}
The BDBCI's state decoder was developed similarly to our prior work in~\cite{wang_benchtop_2019}. Briefly, the decoder preprocessed each window of 32-channel real-time ECoG data by the common average re-referencing and calculating the average power in the two bands. To reduce the dimensionality of the data, $d\in\mathbb{R}^{64}$ (32 channels, 2 bands), and enhance the class-separability, a 1D feature $f\in\mathbb{R}$ was extracted. Formally, $f=T_{\text{LDA}}\Phi_{\text{cPCA}}(d)$, where 
$\Phi_{\text{cPCA}}: \mathbb{R}^{64}\rightarrow \mathbb{R}^m$ is the cPCA mapping with the intermediate dimension, $m$, determined by an eigenvalue criterion~\cite{das_efficient_2009}, and  $T_{\text{LDA}}\in\mathbb{R}^{1\times m}$ is the LDA feature extraction matrix. Due to cPCA's piecewise linear nature, each ECoG data window was projected onto two 1D subspaces: one principally informed by the MOVE state and the other principally informed by the IDLE state.

The parameters of a Bayesian classifier were then estimated offline from the training data, with $P_I(M|f)$ and $P_M(M|f)$ being the posterior probability of walking based on the features in the IDLE and MOVE subspace, respectively. Note that $P_I(I|f) = 1 - P_I(M|f)$ and $P_M(I|f) = 1 - P_M(M|f)$. Subsequently, a real-time data window was transformed into a feature, $f^*$, and the overall posterior probability $P(M|f^*)$ was estimated from the subspace with the highest posterior probability ratio. 

\begin{figure}
    \centering
    \includegraphics[width=0.5\linewidth]{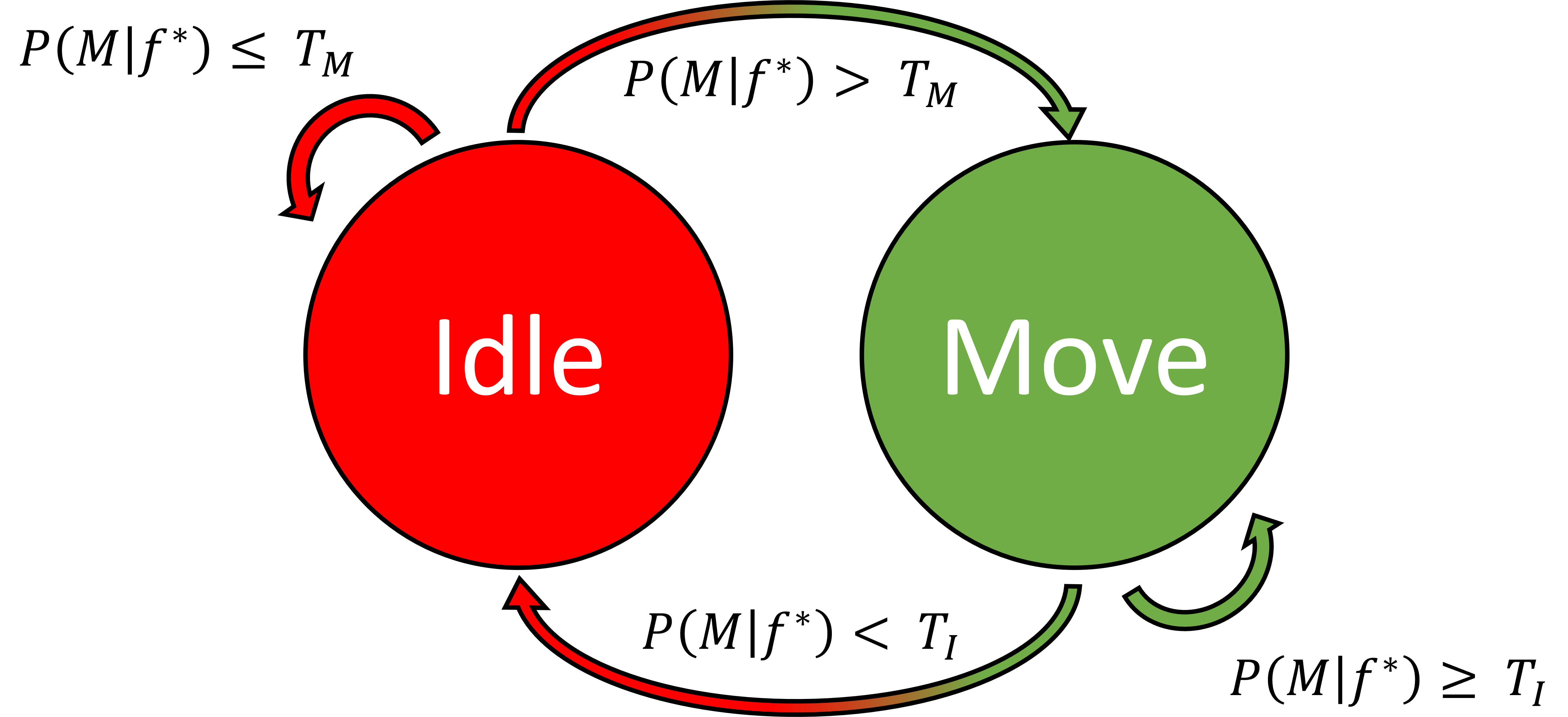}
    \caption{State machine for BDBCI decoder. The state transitions between the Idle and Move state are governed by the state transition rules defined in the figure.}
    \label{fig:decoder_state_machine}
\end{figure}

To determine the decoded state from $P(M|f^*)$, a binary state machine governed the transitions between the ``MOVE" and ``IDLE" states (illustrated in Fig.~\ref{fig:decoder_state_machine}). When $P(M|f^*)$ exceeded the ``MOVE" threshold ($T_M$), the decoded state transitioned from ``IDLE'' to ``MOVE'' state. When $P(M|f^*)<T_I$, the state transitioned from ``MOVE'' to ``IDLE'' state. Finally, when $T_I\le P(M|f^*)\le T_M$, the decoder remained in the  the current state.

\subsection{Decoding Model Optimization} \label{sec:methods_decoder_optimize}
After collecting training data from the subset of ECoG channels selected according to Section~\ref{sec:methods_motormap} and building a preliminary decoding model (Section~\ref{sec:methods:training}), the parameters of the decoder model were further optimized to account for any changes since the motor mapping procedure. 
While the initial subset of channels was chosen from those exhibiting high-$\beta$ and $\gamma$ band modulation (Section~\ref{sec:methods_motormap}) in response to leg movement or those exhibiting motor responses when stimulated (Section~\ref{sec:methods_sensemap}), it was not necessarily the combination that produced the best performance for real-time decoding.

The first step of the optimization utilized a program on the base station computer that evaluated the offline decoding performance of the collected training data by performing a 10-fold cross-validation. 
If the offline results were satisfactory ($\rho>0.85$), an online procedure was performed (Section~\ref{sec:methods_decoderonline}) to evaluate the real-time decoding performance of that model. If either the offline or online decoding performance was unsatisfactory ($\rho<0.85$), the spatial weights (derived from cPCA-LDA matrices) were assessed to see which channels were more/less relevant for decoding (similar process to Fig.~\ref{fig:training_summary}). Those channels with lower-valued spatial weights (i.e., presumed to be less important for decoding) were replaced with other channels identified in the motor mapping or stimulation mapping processes (electrodes exhibiting motor responses from stimulation), and the entire data collection, decoder training, and optimization procedures were repeated. 

The transition thresholds $T_M$ and $T_I$ in the binary state machine (Fig.~\ref{fig:decoder_state_machine}) were also calibrated as follows. By default, $T_M$ was set to 0.6 and $T_I$ was set to 0.4. A calibration online run without RGE control or stimulation was performed while posterior probabilities ($P(M|f^*)$) for each online window were recorded. The 10th percentile of the resulting $P(M|f^*)$ distribution was used to update $T_I$, and the 90th percentile was used to update $T_M$. This calibration process was typically repeated for a minimum of 5 iterations to obtain finalized values for $T_I$ and $T_M$.

\subsection{Decoder Online Procedure} \label{sec:methods_decoderonline}
During online operation, new ECoG data were acquired in 750-ms windows. The decoding procedure described in Section~\ref{sec:methods_decoding} was applied to each windows to obtain decoded states. Decoded states and corresponding instructional cues were transmitted to the base station computer to facilitate subsequent BCI performance analysis. Note that the cues were used purely for results reporting and not as a feature in the decoding model.

\begin{figure}
    \centering
    \includegraphics[width=\linewidth]{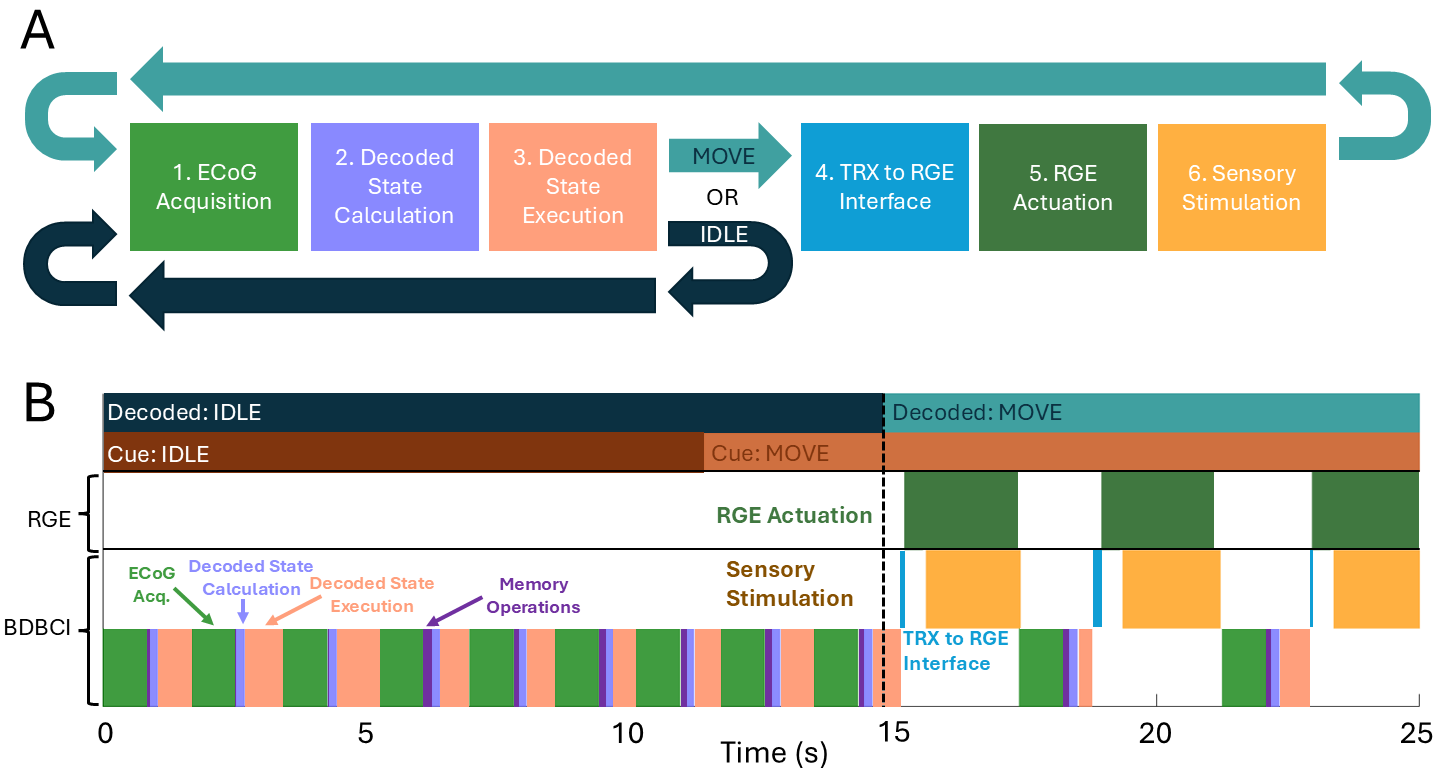}
    \caption{\textbf{A:} High-level state diagram for online motor decoding, RGE control and stimulation operations. (1) BDBCI acquires and stores windows of ECoG data. (2) Decoded state determined using onboard training model. (3) BDBCI acts based on the decoded state. If the IDLE state is decoded, system returns to (1). If the MOVE state is decoded, (4) actuation instructions are transmitted to the RGE Interface, (5) RGE leg swing is actuated, (6) RGE leg swing triggers sensory stimulation in the corresponding leg, and upon completion of the swing, BDBCI returns to (1). \textbf{B:} Example of BDBCI-RGE execution in real time. The first $\sim$13 s demonstrate operations while the decoded state is IDLE. BDBCI executes operations (1) -- (3)  until the decoded state changes to ``MOVE," upon which decoding operations begin to be interleaved with RGE actuation and sensory stimulation (4)--(6).}
    \label{fig:processtimingdiagram}
\end{figure}

In situations where the RGE control was involved, decoded ``MOVE" states were transmitted wirelessly to the RGE Interface for step actuation. As the RGE is incapable of stopping or changing step trajectories mid-step, the decoded state did not need to be updated until a step was completed. By leveraging this constraint, the interleaving of ECoG acquisition/decoding with S1 stimulation, as summarized in Fig.~\ref{fig:processtimingdiagram}, prevented stimulation artifacts from interfering with motor control (c.f. Supplement Section 5 for an example of artifact-free ECoG acquisition during BDBCI-RGE operation). Since it takes $\sim$2.5 s for the RGE to complete a step and $\sim$1 s to acquire and decode one online window, the nominal decoding rate while the RGE was stepping was $\sim$0.29 Hz. Note that the physical stepping speed is the rate-limiting process, and was intentionally set low to permit for safe weight shifting between RGE steps. While the decoded state was ``IDLE", the BDBCI decoding nominal rate is $\sim$1 Hz since the RGE stayed motionless and would not trigger artificial S1 stimulation. 

For all online BDBCI tasks, the decoding performance was evaluated by calculating the lag-optimized Pearson product-moment correlation coefficient between the cues and the decoded states that were streamed to the base station computer during the online procedure. The correlations quantified the decoding model's efficacy, as well as the subject's ability to operate the BDBCI with all available visual and artificial sensory feedback.

\subsection{BDBCI-RGE Operation} \label{sec:methods_bdbci_task}

BDBCI-RGE operation involved control of the RGE using ECoG signals underlying bilateral stepping-like movements, while visual and artificial sensory feedback were provided with each RGE step. 
We used the motor decoding model generation (Section~\ref{sec:methods_decoding}) and optimization (Section~\ref{sec:methods_decoder_optimize}) to enable neural walking control. Stimulation parameters for left and right leg percepts were determined as described above in Section~\ref{sec:methods_sensemap}), and were triggered by the left and right RGE leg swings. For safety reasons, an experimenter, rather than the subject, wore the RGE, and the subject assumed BDBCI control of the RGE from the hospital bed via wireless connection. Prior to the BDBCI-RGE Task, two control experiments were performed (Section~\ref{sec:bdbci}), each turning off RGE control or S1 stimulation to verify that neither visual feedback of the RGE nor S1 DCES drove the motor decoding.
Real-time control of the BDBCI-RGE system was performed as described in Section~\ref{sec:methods_decoderonline}. As BDBCI-RGE operation utilizes stimulation, the interleaved decoding mode was used (Fig. \ref{fig:processtimingdiagram}). Specifically, the RGE Interface triggered the BDBCI to elicit the percept of the corresponding leg for the duration of the leg swing at the onset of each RGE leg swing. In addition, ECoG acquisition was suspended during the RGE leg swing to prevent interference from electrical stimulation artifacts in the ECoG signal from compromising the decoding functionality. Once the leg swing ended (as detected by the RGE Interface), artificial sensory stimulation stopped and ECoG acquisition resumed. The decoded state of the newly acquired ECoG window determined whether the RGE continued stepping or transitions to standing/idle.

\subsection{Kinematic Acquisition} \label{sec:methods_xsens}
The kinematics of the experimenter wearing the RGE were captured using the Xsens 3D motion capture system (Movella Inc., San Jose, CA, USA). These sensors were placed on the lower extremity and captured the RGE-mediated trajectory. These data were streamed wirelessly to the base station computer and recorded at 100 Hz. Angular velocities were obtained and aligned to BDBCI log entries for data visualization.

\backmatter

\section*{Declarations}

\begin{itemize}
\item Funding: This work was supported by the National Science Foundation (NSF) Grant number 1646275
\end{itemize}

\subsection*{Ethics Statement}
All studies were performed with the approval of the the Institutional Review Boards of the University of California, Irvine and Rancho Los Amigos National Rehabilitation Center.
\bibliography{BDBCIArXiv}

\end{document}


\title[Article Title]{Supplemental Material}

\section{Subject Clinical Details}

The main subject in this study was a known medication refractory epilepsy patient with onset at age of ten. Seizure semiology is described as starring, sense of intense fear, screaming, followed by full body stiffening and eye fluttering. Seizure frequency was typically twice a month, but can be up to 5 per month.
Extensive prior workup included an MRI scan showing left hippocampal sclerosis, and a PET scan showing hypermetabolism in the left mesial temporal lobe. These findings, which suggested seizure origin in the left mesial temporal lobe, did not align with EEG recordings demonstrating bifrontal onset seizure onset during prior Phase 1 epilepsy monitoring unit evaluation. After discussion at an epilepsy conference, it was recommended that a Phase 2 epilepsy monitoring unit evaluation would be necessary to better characterize seizures,  untangle imaging and physiological discrepancy, and determine if patient was an epilepsy surgery candidate. 

The patient consented to the Phase 2 epilepsy monitoring evaluation, which involved a bifrontal craniotomy for placement of bilateral interhemispheric and frontal convexity subdural strip and grid electrodes, as well as left temporal craniotomy for placement of subdural sub-temporal strip electrodes under general anesthesia. The patient was intubated and vascular access lines were established.  The patient's head was shaved and securely positioned to the right in a cranial stabilization device (Mayfield head clamp, Integra, Princeton, NJ). The patient's head was registered to a surgical navigation system (STEALTH Neuro-navigation System, Medtronic, Minneapolis, MN). The surgical site was prepared and draped to establish a sterile surgical field. The midline scalp incision was made and a musculocutaneous flap was turned and retracted anteriorly. A midline frontal craniotomy was then performed using 6 burr holes. The dura to the left of the midline was opened and an interhemispheric 8$\times$8 high density ECoG grid was placed under visualization with a high-power operating microscope so as to preserve all bridging veins. This was followed by placement of a 2$\times$6 anterior frontal and 1$\times$6 lateral frontal ECoG electrode grid/strip. Subsequently, using similar meticulous microsurgical techniques to preserve all bridging veins, the dura to the right of midline was opened and an interhemispheric 8$\times$8 high density ECoG grid was placed, followed by a 2$\times$6 anterior frontal and 1$\times$6 lateral ECoG frontal grid/strip. 
All electrode arrays and strips were secured to the dura and connected to the clinical bioamplifier system (Natus\textsuperscript{\textregistered} Quantum$^\textnormal{TM}$, Natus Medical Incorporated, Pleasanton, CA, USA). ECoG signal recordings were verified by the epileptologist to be acceptable. Meticulous hemostasis was achieved, the bone flap was replaced, and all electrode leads were tunneled out and secured to the skin in a watertight fashion.  
Next, the left temporal craniotomy was performed, with two 1$\times$6 ECoG grid strips placed in the sub-temporal region.
All ECoG leads were connected to the clinical bioamplifier and ECoG signals were verified to be acceptable  before they were tunneled out and secured to the skin in a watertight fashion.  
Intraoperative imaging (O-arm, Medtronic, Minneapolis, MN) was performed to confirm all electrode placements. 
Finally, the wound was closed and the patient's head was dressed. Anesthesia was reversed and the patient was extubated without difficulty and taken to the recovery room in stable condition.

\section{Sensory Mapping Responses}
\begin{figure}[!htpb]
    \centering
    \includegraphics[width=0.4725\linewidth]{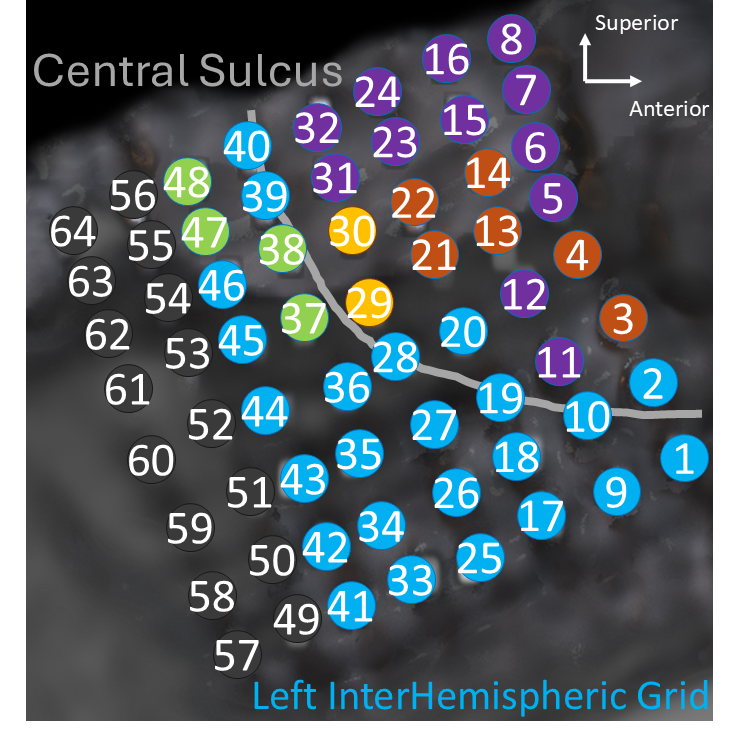}
    \includegraphics[width=0.485\linewidth]{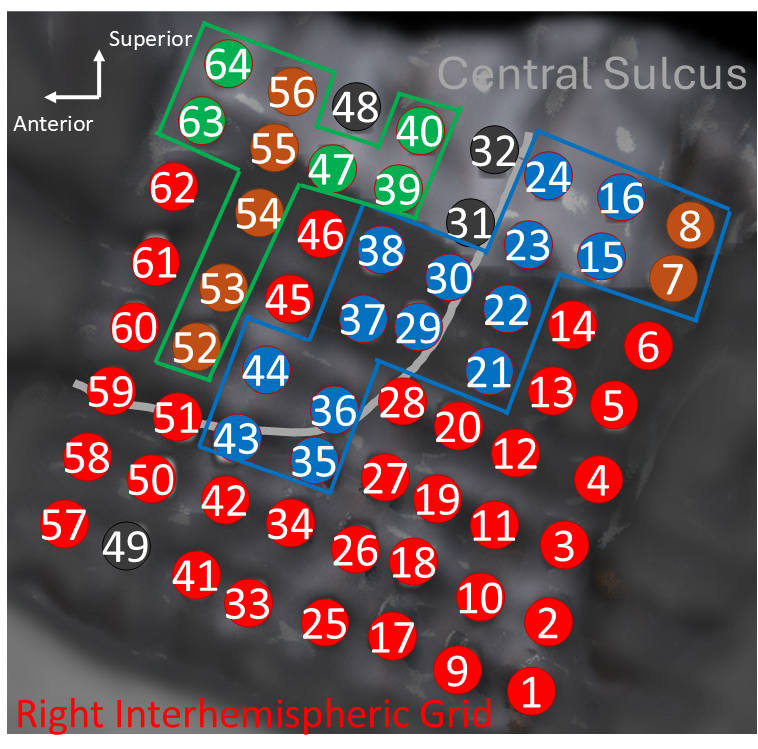}
    \caption{\textbf{Left:} Clinical cortical mapping responses mapped to ECoG electrode locations on the left inter-hemispheric ECoG grid. Purple: right leg motor response. Orange: right leg and arm motor response. Yellow: right foot sensory response. Green: non-specific whole-body sensation. Cyan: stimulated but no response. Grey: not stimulated. For a full documenting of elicited responses, see Table~\ref{tab:natus_stim_mapping}. \textbf{Right:} Clinical cortical mapping responses mapped to ECoG electrode locations on the right inter-hemispheric ECoG grid. Blue: Left leg motor response. Green: Bilateral leg motor response. Orange: sensory response reported. Red: stimulated but no response. Grey: not stimulated. Note that some electrodes had both sensory and motor responses (e.g. L7-8, L52-56). For a full documenting of elicited responses, see Table~\ref{tab:natus_stim_mapping}. }
    \label{fig:clinicalmapping}
\end{figure}

\begin{figure}[!htpb]
    \centering
    \includegraphics[width=0.4725\linewidth]{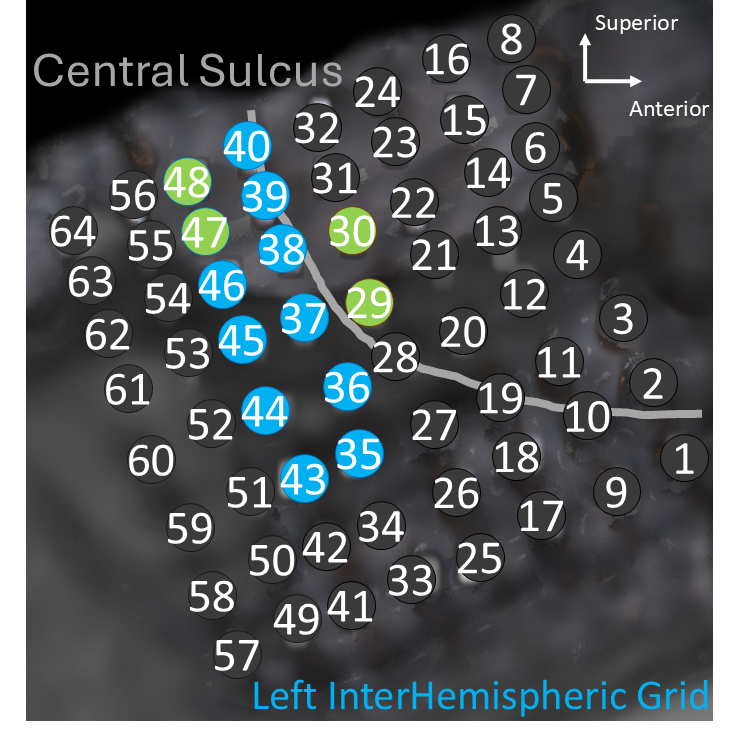}
    \includegraphics[width=0.485\linewidth]{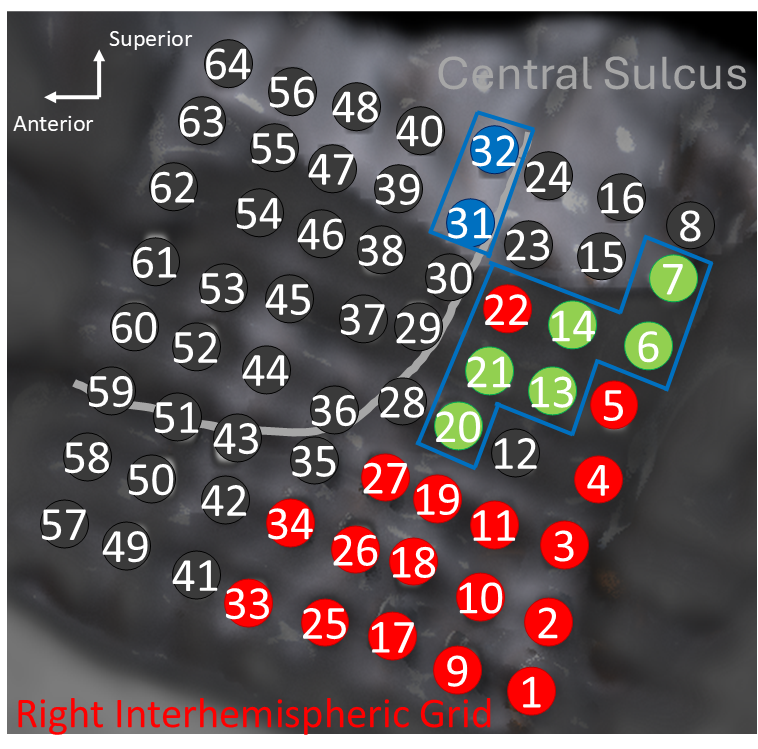}
    \caption{\textbf{Left:} BDBCI cortical mapping responses mapped to ECoG electrode locations on the left inter-hemispheric ECoG grid. Purple: right leg motor response. Green: sensory response reported. Cyan: stimulated but no response. Grey: not stimulated. For a full documenting of elicited responses, see Table~\ref{tab:bdbci_stimnotes}. \textbf{Right:} BDBCI cortical mapping responses mapped to ECoG electrode locations on the right inter-hemispheric ECoG grid. Blue: Left leg motor response. Green: sensory response reported. Red: stimulated but no response. Grey: not stimulated. Note that some electrodes had both sensory and motor responses (e.g. L7-8, L52-56). For a full documenting of elicited responses, see Table~\ref{tab:bdbci_stimnotes}. }
    \label{fig:bdbcimapping}
\end{figure}

The responses for the sensory mapping using the clinical cortical stimulator and the BDBCI are shown in the following figures and tables. The stimulation responses for the left and right grids using the clinical cortical stimulator are shown in Fig. \ref{fig:clinicalmapping}.
A tabulated list of all stimulation parameters eliciting responses using the clinical cortical stimulator are documented in Table~\ref{tab:natus_stim_mapping}. Similarly, Fig.~\ref{fig:bdbcimapping} maps the high-level BDBCI stimulation mapping responses to the ECoG electrode locations, while Table~\ref{tab:bdbci_stimnotes} documents all responses elicited.

\newpage

\begin{longtable}{p{0.11\linewidth}p{0.075\linewidth}p{0.14\linewidth}p{0.6\linewidth}}
\caption{Clinical Cortical Mapping Responses} \\
\toprule
Channel & Stim. Amp. (mA)                   & Response Type & Description \\
\midrule
R7-R8       & 2                 & Sensory       & Left sole, pins and needles                                                                                                                                                    \\
                         & 8                             & Motor         & left foot dorsiflexion                                                                                                                                              \\
R15-R16      & 4                             & Motor         & Left foot dorsiflexion                                                                                                                                              \\
R23-R24      & 4                             & Motor         & Left foot dorsiflexion                                                                                                                                              \\
R21-R22      & 2                        & Motor         & Left knee flexion, left foot dorsiflexion                                                                                                                           \\
R35-R36      & 10                        & Motor         & Left foot inversion, left hip abduction                                                                                                                             \\
R37-R38      & 6                         & Motor         & Left foot dorsiflexion                                                                                                                                              \\
R39-R40      & 6                    & Motor         & Bilateral feet, knee extension, hip abduction                                                                                                                            \\
R43-R44      & 6                         & Motor         & Left foot inversion, left hip abduction                                                                                                                             \\
R46-R47      & 4                    & Motor         & Bilateral foot plantarflexion, hip abduction, knee extension                                                                                                       \\
R52-R53      & 6                    & Motor         & Bilateral foot plantarflexion, hip abduction, knee extension, felt pain as though she stepped on an object                                                           \\
R53-R54      & 6                    & Sensory and motor  & Bilateral foot plantarflexion, hip abduction, knee extension, felt pain as though she stepped on an object. Only felt in the left leg, like someone is   pulling her \\
R55-R56     & 8                    & Sensory and motor  & Bilateral foot plantarflexion, hip abduction, knee extension, felt pain as though she stepped on an object                                                                        \\
R59-R60      & 6                         & Motor         & Left foot inversion, right foot eversion, bilateral hip abduction                                                                                             \\
R63-R64      & 4             & Motor         & Left leg extension, left foot plantarflexion, right knee flexion                                                                                                    \\
R63-R64      & 6                    & Motor         & Right leg extension and right hip adduction/inversion (inward). Left foot   abduction, left foot inversion 
\\
\midrule
L7-L8       & 6                       & Motor         & Right foot dorsiflexion, right knee flexion                                                                                                                         \\
L5-L6       & 6                       & Motor         & Strong right knee flexion, right foot dorsiflexion, right toes plantarflexion                                                                                                    \\
L3-L4       & 4                         & Motor         & Right arm extension, right foot inversion                                                                                                                           \\
L15-L16      & 4                            & Motor         & Right foot slight inversion                                                                                                                                         \\
L13-L14      & 6                       & Motor         & Right foot dorsiflexion, right knee flexion                                                                                                                      \\
L11-L12      & 8                            & Motor         & Right foot dorsiflexion and inversion                                                                                                                                   \\
L23-L24      & 6                            & Motor         & Right foot dorsiflexion, right toes plantarflexion                                                                                                                               \\
L21-L22      & 6                   & Motor         & Right hip flexion, right foot eversion, right elbow extension \\
L31-L32      & 10                           & Motor         & Right foot plantarflexion, right toes plantarflexion \\
L29-L30      & 10                           & Sensory       & Pins and needles at the right-outer bottom of the right foot                                                                                                        \\
L37-L38      & 10                           & Sensory       & Whole body sensation that subject could not describe                                                                                                                                                   \\
L47-L48      & 6                            & Sensory       & Whole body sensation that subject could not describe                                                                                                                                                   \\                                                                                                              
\bottomrule
\label{tab:natus_stim_mapping}
\end{longtable}

\begin{table}[!htpb]
\label{tab:bdbci_stimnotes}
\caption{BDBCI Cortical Mapping Responses}
\begin{tabular}{p{0.1\linewidth}p{0.05\linewidth}p{0.05\linewidth}p{0.1\linewidth}p{0.70\linewidth}}
\toprule
Channel & Freq. (Hz) & Curr. (mA)                                          & Response Type & Description                                                                                                                          \\
\midrule
L29-L30      & 100       & 7.48                                                & Sensory       & Pins and needles on the right heel                                                                                                              \\
L29-L30      & 100       & 7.85                                  & Sensory       & Pins and needles migrating from right sole to right calf                                                                                             \\
L47-L48      & 100       & 4.71                                                & Sensory       & Electric Currents moving upward from right heel                                                                                               \\
L47-L48      & 200       & 4.95          & Sensory       & Electric Currents moving upward from right foot to right shoulder                                                                                          \\
\midrule
R6-R7       & 100       & 3.37                               & Sensory       & Pins and needles on lateral side of left calf                                                                                                              \\
R6-R7       & 150       & 7.69                                                 & Sensory and Motor  & Left toe and foot dorsiflexion, pins and needles in the left foot                                                            \\
R13-R14      & 100       & 2.25                                             & Sensory and Motor  & Left foot dorsiflexion; difficult to describe sensation; starts from bottom and moves up to calf but not knee \\
R13-R14      & 200       & 3.8                                                  & Motor         & Left toes dorsiflexion, then foot dorsiflexion                                                                                                   \\
R20-R21      & 100       & 6.07                                            & Motor         & left foot dorsiflexion, neck extension on the left side                                                                                           \\
R21-R22      & 100       & 4.75                                                  & Motor         & Bilateral feet dorsiflexion                                                                                                        \\
R31-R32      & 100       & 1.55                                             & Motor         & Left hip flexion, foot, and toes dorsiflexion                                                                              \\
\bottomrule
\end{tabular}
\end{table}

\newpage

\section{Sensory Task Results}
\begin{figure}[!htpb]
    \centering
    \includegraphics[width=\linewidth]{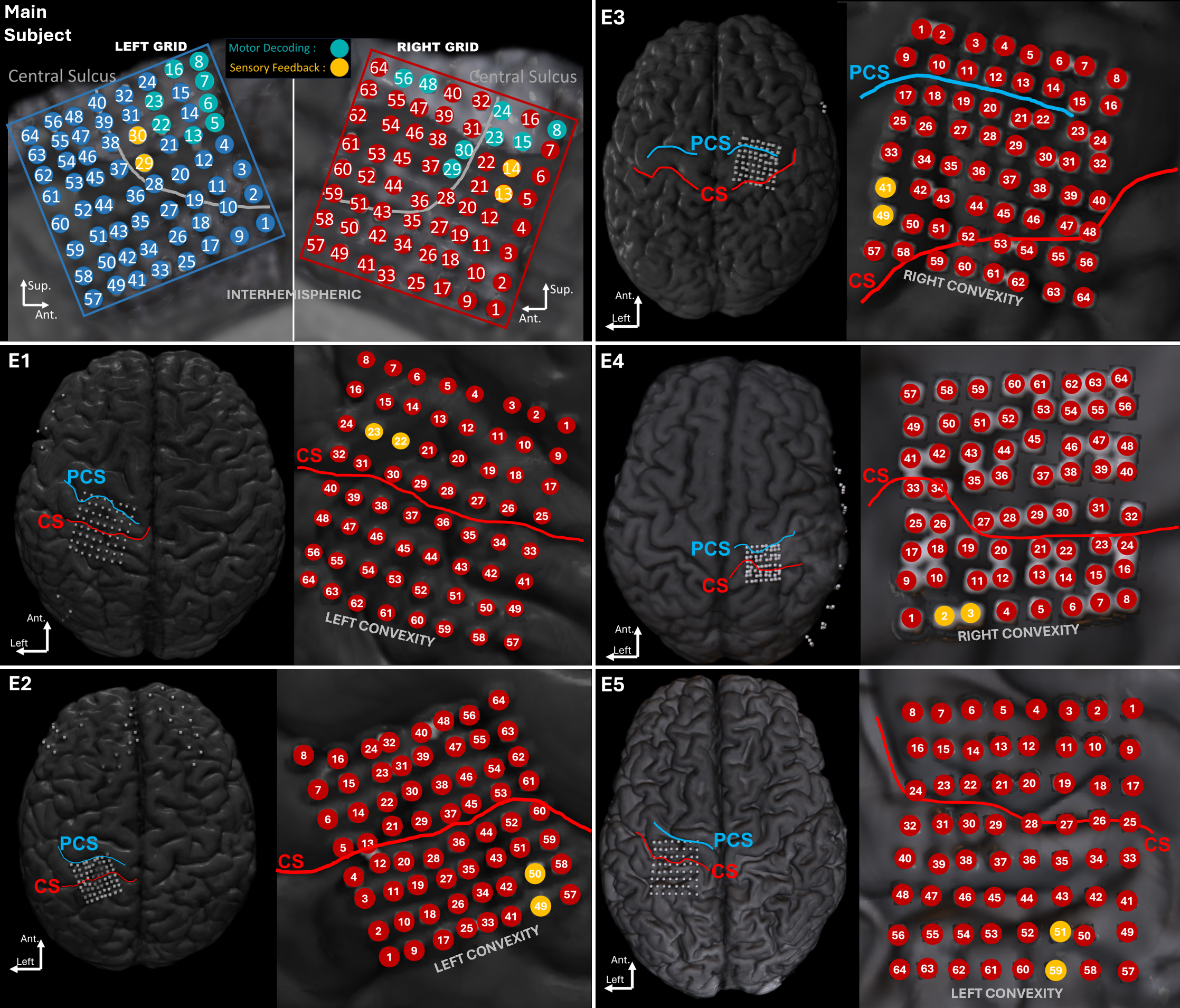}
    \caption{CT-MR co-registered images depicting ECoG grid locations for each subject (Main Subject and Extra Subjects 1 - 5). Electrodes marked in yellow were for sensory stimulation during the step counting task, as well as for blind sensory testing for Main Subject and Extra Subject 2.}
    \label{fig:coreg_collage}
\end{figure}

The results for sensory tasks performed by the subject described in the main manuscript (Main Subject) have been documented here. Sensory task performances by other subjects (Extra Subject 1 - 5, all subjects provided informed consent to participate in this study.) that have previously worked with our group have also been included. For step counting results, see Table~\ref{tab:step_count_trials_table}, which shows the counted steps per-trial for each of the subjects. The number of correctly counted trials for each subject are tabulated in Table~\ref{tab:all_patient_step_trials}. Over all step-counting trials, no subject had an error larger than $\pm$2 steps counted. None of the counting results were likely to have been achieved by random guessing, as indicated by the empirical p-values for each of the subjects ($p < 10^{-6}$). For blind sensory test results, see Table~\ref{tab:blind_sensory_results}. Note that only the Main Subject and  Extra Subject 2 performed the blind sensory test. Additionally, Extra Subject 2 only had a lateral convexity implant, so only right leg percepts are compared against the null sensation. Neither of these subjects' results were likely to have been achieved by random guessing (empirical p-values, $p < 10^{-6}$). For added context, the co-registration images of each Extra Subject, as well as the electrodes used for sensory feedback are included in Fig.~\ref{fig:coreg_collage}. 

\begin{longtable}{lllll}
\caption{Step Counting Experiment Raw Results for Left and Right Leg Sensation} \\
    \toprule
    \multicolumn{5}{c}{Main Subject } \\
    \midrule
    Trial & True Step (R) & Counted Steps (R) & True Step (L) & Counted Steps (L) \\
    \midrule
    1     & 3             & 3                      & 7             & 7                      \\
    2     & 2             & 2                      & 2             & 2                      \\
    3     & 8             & 8                      & 6             & 6                      \\
    4     & 2             & 2                      & 8             & 7                      \\
    5     & 4             & 4                      & 3             & 3                      \\
    6     & 7             & 7                      & 5             & 5                      \\
    7     & 3             & 3                      & 6             & 6                      \\
    8     & 7             & 7                      & 5             & 5                      \\
    9     & 4             & 4                      & 4             & 4                      \\
    10    & 8             & 8                      & 3             & 3                      \\
    11    & 5             & 5                      & 4             & 4                      \\
    12    & 5             & 5                      & 2             & 2                      \\
    13    & 6             & 6                      & 9             & 7                      \\
    14    & 6             & 6                      & 7             & 7                     \\
\midrule
    empirical p-value & $<$ 10$^{-6}$ & & & \\
\midrule
\multicolumn{5}{c}{Extra Subject 1 (F, 21)} \\
\midrule
Trial & True Count & Counted Steps &  & \\
\midrule
1     & 3             & 3                      &              &                       \\
2     & 2             & 2                      &              &                       \\
3     & 8             & 8                      &              &                       \\
4     & 2             & 2                      &              &                       \\
5     & 4             & 4                      &              &                       \\
6     & 7             & 7                      &              &                       \\
7     & 3             & 3                      &              &                       \\
8     & 7             & 7                      &              &                       \\
9     & 4             & 4                      &              &                       \\
10    & 9             & 8                      &              &                       \\
11    & 5             & 6                      &              &                       \\
12    & 5             & 5                      &              &                       \\
13    & 6             & 7                      &              &                       \\
14    & 6             & 6                      &              &                      \\
\midrule
    empirical p-value & $<$ 10$^{-6}$ & & & \\
\midrule
\multicolumn{5}{c}{Extra Subject 2 (F, 42)} \\
\midrule
Trial & True Count & Counted Steps &  & \\
\midrule
1     & 3             & 3                      &              &                       \\
2     & 2             & 2                      &              &                       \\
3     & 8             & 8                      &              &                       \\
4     & 2             & 2                      &              &                       \\
5     & 4             & 4                      &              &                       \\
6     & 7             & 7                      &              &                       \\
7     & 3             & 2                      &              &                       \\
8     & 7             & 6                      &              &                       \\
9     & 4             & 3                      &              &                       \\
10    & 9             & 7                      &              &                       \\
11    & 5             & 5                      &              &                       \\
12    & 5             & 5                      &              &                       \\
13    & 6             & 7                      &              &                       \\
14    & 6             & 4                      &              &                      \\
\midrule
    empirical p-value & $<$ 10$^{-6}$ & & & \\
\midrule
\multicolumn{5}{c}{Extra Subject 3 (F, 32)} \\
\midrule
Trial & True Count & Counted Steps &  & \\
\midrule
1     & 3             & 2                      &              &                       \\
2     & 2             & 2                      &              &                       \\
3     & 8             & 7                      &              &                       \\
4     & 2             & 2                      &              &                       \\
5     & 4             & 5                      &              &                       \\
6     & 7             & 7                      &              &                       \\
7     & 3             & 3                      &              &                       \\
8     & 7             & 7                      &              &                       \\
9     & 4             & 3                      &              &                       \\
10    & 8             & 8                      &              &                       \\
11    & 5             & 4                      &              &                       \\
12    & 5             & 5                      &              &                       \\
13    & 6             & 6                      &              &                       \\
14    & 6             & 4                      &              &                      \\
\midrule
    empirical p-value & $<$ 10$^{-6}$ & & & \\
\midrule
\multicolumn{5}{c}{Extra Subject 4 (F, 47) } \\
\midrule
Trial & True Count & Counted Sensations &  & \\
\midrule
1     & 3             & 3                      &              &                       \\
2     & 2             & 2                      &              &                       \\
3     & 8             & 8                      &              &                       \\
4     & 2             & 2                      &              &                       \\
5     & 4             & 4                      &              &                       \\
6     & 7             & 7                      &              &                       \\
7     & 3             & 3                      &              &                       \\
8     & 7             & 7                      &              &                       \\
9     & 4             & 4                      &              &                       \\
10    & 9             & 9                      &              &                       \\
11    & 5             & 5                      &              &                       \\
12    & 5             & 5                      &              &                       \\
13    & 6             & 6                      &              &                       \\
14    & 6             & 6                      &              &                      \\
\midrule
    empirical p-value & $<$ 10$^{-6}$ & & & \\
\midrule
\multicolumn{5}{c}{Extra Subject 5 (M, 22)} \\
\midrule
Trial & True Count & Counted Steps &  & \\
\midrule
1     & 3             & 3                      &              &                       \\
2     & 2             & 2                      &              &                       \\
3     & 8             & 8                      &              &                       \\
4     & 2             & 2                      &              &                       \\
5     & 4             & 4                      &              &                       \\
6     & 7             & 6                      &              &                       \\
7     & 3             & 3                      &              &                       \\
8     & 7             & 7                      &              &                       \\
9     & 4             & 4                      &              &                       \\
10    & 9             & 9                      &              &                       \\
11    & 5             & 5                      &              &                       \\
12    & 5             & 5                      &              &                       \\
13    & 6             & 6                      &              &                       \\
14    & 6             & 6                      &              &                      \\
\midrule
    empirical p-value & $<$ 10$^{-6}$ & & & \\
\midrule
\bottomrule
\label{tab:step_count_trials_table} \\
\end{longtable}

\begin{table}[!htpb]
    \caption{Step counting results: Number of correctly counted trials (out of 14).}
    \centering
    \begin{tabular}{ccc}
        \toprule
         Subject &  Number Trials Correct (of 14) & Largest Count Error \\
         \midrule
         Main Subject (R) & 14  & 0 \\
         Main Subject (L) & 12 &  +2\\
         Extra Subject 1 & 11 &  +1\\
         Extra Subject 2 & 8 &  -2\\
         Extra Subject 3 & 8 &  -2\\
         Extra Subject 4 & 14 &  0\\
         Extra Subject 5 & 13 &  -1\\
         \midrule
         Average$\pm$1SD & 11.4$\pm$2.6 & -0.3$\pm$1.5 \\
         \bottomrule
    \end{tabular}
    \label{tab:all_patient_step_trials}
\end{table}

\newpage

\newpage

\begin{table}[!htpb]
\caption{Blind Sensory Test Results}
\begin{tabular}{llll}
\toprule
\multicolumn{4}{l}{Main Subject} \\

\midrule
& Null & Right & Left \\
\midrule
Null  & 25   & 0    & 0     \\
Right  & 1    & 24   & 0     \\
Left & 4    & 0    & 21   \\
\midrule
\multicolumn{4}{l}{Extra Subject 2} \\
\midrule
 & Null & Right & \\
 \midrule
Null  & 25   & 0    &      \\
Right  & 3    & 21   &      \\
\bottomrule
\end{tabular}
\label{tab:blind_sensory_results}
\end{table}

\newpage
\section{BCI Decoding Performances}
Performances of the various decoding tasks described in the manuscript are documented in Table.~\ref{tab:online_decoding_results}. 

\begin{table}[!htpb]
\caption{Real-Time decoding results for BCI tasks}
\begin{tabular}{lllll}
\toprule
                       & Max Corr      & Lag (s) & $T_I$   & $T_M$   \\
\midrule
\multicolumn{5}{c}{Familiarization Task (BCI only)} \\
\midrule
Calibration Run        & 0.816         & 2       & 0.40  & 0.60  \\
Run 1                  & 0.704         & 1       & 0.94 & 0.98 \\
Run 2                  & 0.805         & 3       & 0.50  & 0.97 \\
Run 3                  & 0.915         & 2       & 0.38 & 0.84 \\
Run 4                  & 0.919         & 2       & 0.18 & 0.84 \\
Run 5                  & 0.943         & 2       & 0.10  & 0.86 \\
Avg + 1SD              & 0.888 $\pm$ 0.056 & 2.2 $\pm$ 0.4 &      &      \\
\midrule

\multicolumn{5}{c}{Day 1 of BDBCI-RGE} \\
\midrule
Run 1                  & 0.865         & 3 & 0.02     & 0.5\\
Run 2                  & 0.913         & 3 & 0.02     & 0.5      \\
Run 3                  & 0.911         & 4 & 0.02     & 0.5      \\
Run 4                  & 0.876         & 4 & 0.02     & 0.5      \\
Run 5                  & 0.888         & 3 & 0.02     & 0.5      \\
Avg + 1SD              & 0.891 $\pm$ 0.021 & 3.4 $\pm$ 0.5      &      \\
\midrule
\multicolumn{5}{c}{Day 2 of BDBCI-RGE} \\
\midrule
Run 1                  & 0.967         & 3 & 0.02     & 0.95     \\
Run 2                  & 0.906         & 4 & 0.02     & 0.95      \\
Run 3                  & 0.963         & 4 & 0.02     & 0.95      \\
Run 4                  & 0.913         & 3 & 0.02     & 0.95      \\
Run 5                  & 0.969         & 4 & 0.02     & 0.95      \\
Avg + 1SD              & 0.944 $\pm$ 0.028 & 3.6 $\pm$ 0.5     &     \\
\midrule
\multicolumn{5}{c}{BCI-RGE Task (Control)} \\
\midrule
Run 1                  & 0.859         & 4 & 0.07     & 0.86     \\
Run 2                  & 0.951         & 4 & 0.07     & 0.86      \\
Run 3                  & 0.927         & 2 & 0.07     & 0.86      \\
Run 4                  & 0.923         & 4 & 0.07     & 0.86      \\
Run 5                  & 0.940         & 4 & 0.07     & 0.86      \\
Avg + 1SD              & 0.932 $\pm$ 0.013 & 3.6 $\pm$ 0.9     &      \\
\midrule
\multicolumn{5}{c}{BCI-Stimulation (Control)} \\
\midrule
Run 1                  & 0.951         & 3 & 0.02     & 0.5     \\
Run 2                  & 0.911         & 4 & 0.02     & 0.5      \\
Run 3                  & 0.968         & 4 & 0.02     & 0.5      \\
Run 4                  & 0.950         & 4 & 0.02     & 0.5      \\
Run 5                  & 0.919         & 4 & 0.02     & 0.5      \\
Avg + 1SD              & 0.940 $\pm$ 0.024 & 3.8 $\pm$ 0.4&      &      \\
\bottomrule
\end{tabular}
\label{tab:online_decoding_results}
\end{table}

\newpage
\section{Example of Neural Data Acquisition with Cortical Electrostimulation}

\begin{figure}[!htpb]
    \centering
    \includegraphics[width=\linewidth]{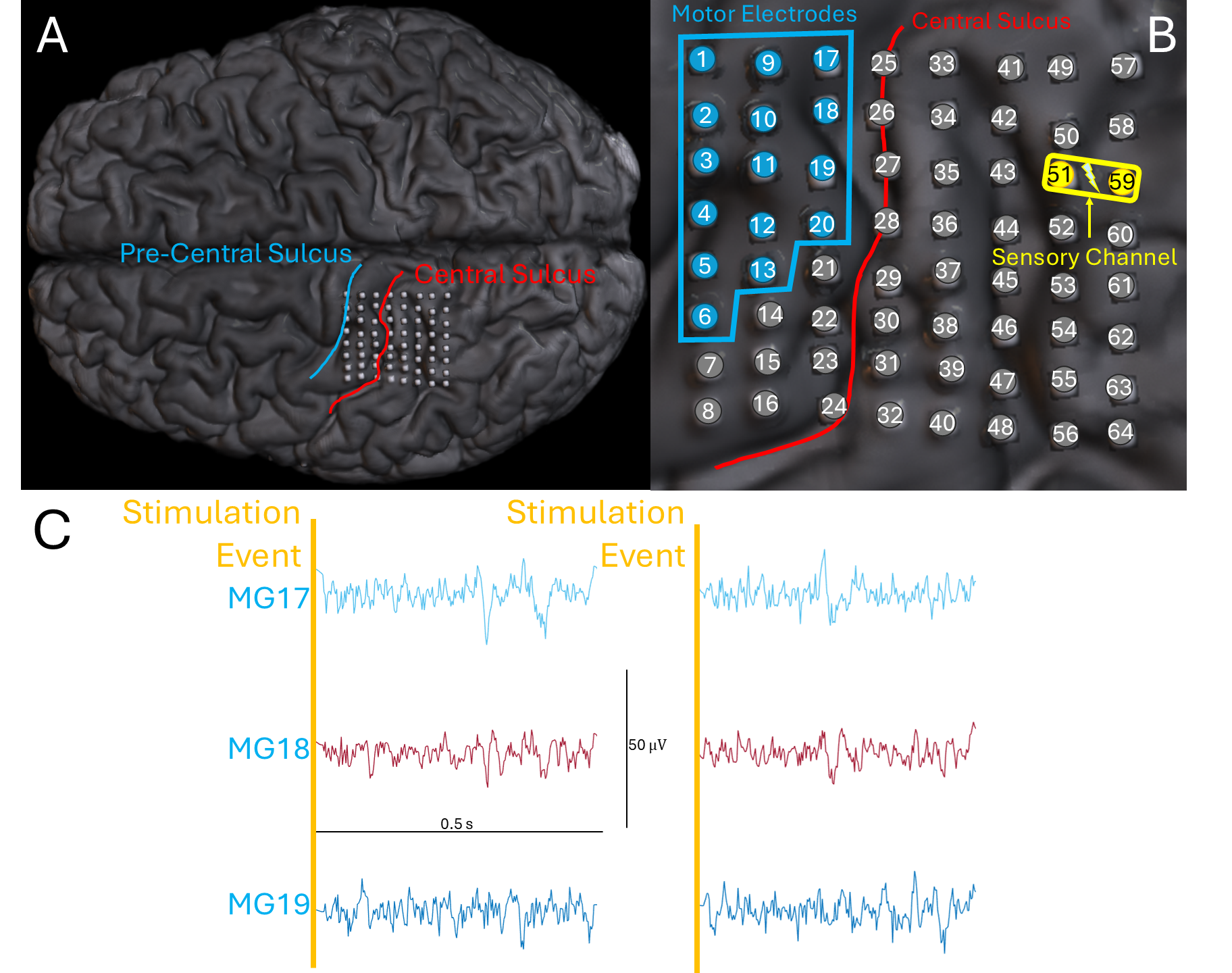}
    \caption{ECoG signals acquired during BDBCI-RGE task (Extra Subject 2). \textbf{A:} The placement of the ECoG grid for this subject. \textbf{B:} Motor decoding and stimulation channels. \textbf{C:} ECoG signals acquired between stimulation events do not exhibit any electrical artifacts, even on those electrodes closest to the stimulation channel ($\sim$16 mm away).}
    \label{fig:online_interleave_example}
\end{figure}

Fig.~\ref{fig:online_interleave_example} shows representative example of ECoG signals acquired during a BDBCI-RGE task from a past patient [18] (Extra Subject 2 from previous sections). ECoG signals immediately proceeding sensory electrocortical stimulation triggered by RGE movement contained no stimulation artifacts. This indicates that the interleaving methodology is successful in preventing electrical interference from disrupting online motor decoding. 

